\documentclass[aps,prd,preprint,superscriptaddress,showpacs,floatfix,nobibnotes]{revtex4}

\usepackage{latexsym}
\usepackage{amsmath}
\usepackage{amssymb}
\usepackage{graphicx}
\usepackage{longtable}

\usepackage{bm}
\usepackage{epsfig}
\usepackage{citesort}

\begin{document}
\title{Chiral Magnetic Effect and QCD Phase Transitions with Effective Models}

\author{Wei-jie Fu}
\email[]{wjfu@itp.ac.cn} \affiliation{Kavli Institute for
Theoretical Physics China (KITPC), Key Laboratory of Frontiers in
Theoretical Physics, Institute of Theoretical Physics, Chinese
Academy of Science, Beijing 100190, China}

\author{Yu-xin Liu}
\email[]{yxliu@pku.edu.cn} \affiliation{Department of Physics and
State Key Laboratory of Nuclear Physics and Technology, Peking
University, Beijing 100871, China} \affiliation{Center of
Theoretical Nuclear Physics, National Laboratory of Heavy Ion
Accelerator, Lanzhou 730000, China}

\author{Yue-liang Wu}
\email[]{ylwu@itp.ac.cn} \affiliation{Kavli Institute for
Theoretical Physics China (KITPC), Key Laboratory of Frontiers in
Theoretical Physics, Institute of Theoretical Physics, Chinese
Academy of Science, Beijing 100190, China}

\date{\today}

\begin{abstract}
We study the influence of the chiral phase transition on the chiral
magnetic effect. The chiral electric current density along the
magnetic field, the electric charge difference between on each side
of the reaction plane, and the azimuthal charged-particle
correlations as functions of the temperature during the QCD phase
transitions are calculated. It is found that with the decrease of
the temperature, the chiral electric current density, the electric
charge difference, and the azimuthal charged-particle correlations
all get a sudden suppression at the critical temperature of the
chiral phase transition, because the large quark constituent mass in
the chiral symmetry broken phase quite suppresses the axial anomaly
and the chiral magnetic effect. We suggest that the azimuthal
charged-particle correlations (including the correlators divided by
the total multiplicity of produced charged particles which are used
in current experiments and another kind of correlators not divided
by the total multiplicity) can be employed to identify the
occurrence of the QCD phase transitions in RHIC energy scan
experiments.
\end{abstract}

\pacs{25.75.Nq  %Quark deconfinement, quark-gluon plasma production, and phase transitions
      11.30.Er, %Charge conjugation, parity, time reversal, and other discrete symmetries
      11.30.Rd, %Chiral symmetry
      11.30.Qc, %Spontaneous and radiative symmetry breaking
      }

\maketitle

\section{Introduction}
\vspace{5pt}

The phase transitions of quantum chromodynamics (QCD), for example
the phase transition of the chiral symmetry restoration and the
deconfinement phase transition, have attracted lots of attentions in
recent years. It is expected that these phase transitions occur and
the deconfined quark gluon plasma (QGP) is formed in
ultrarelativistic heavy-ion
collisions~\cite{Arsene2005,Back2005,Adams2005,Adcox2005,Shuryak2004}
(for example the current experiments at the Relativistic Heavy Ion
Collider (RHIC) and the Large Hadron Collider (LHC)) and in the
interior of neutron stars~\cite{Weber2005,Alford2008,Fu2008b}.
Furthermore, studying the QCD phase transitions is also an
elementary problem in strong interaction physics.

Recently, The STAR Collaboration at RHIC report their measurements
of azimuthal charged-particle correlations near center-of-mass
rapidity in Au + Au and Cu + Cu collisions at
$\sqrt{s_{NN}}=200\,\mathrm{GeV}$. They find a significant signal
consistent with the charge separation of quarks along the system's
orbital angular momentum axis~\cite{Abelev2009a,Abelev2009b}. The
observed charge separation indicates that parity-odd domains, where
the parity ($\mathcal{P}$) symmetry is locally violated, might be
created during the relativistic heavy-ion
collisions~\cite{Kharzeev2006,Kharzeev2008}.

QCD is an SU(3) Yang-Mills gauge theory coupled with quarks. The
gauge field can have nontrivial configurations which can be
characterized by a topological invariant, the winding number
$Q_{w}$~\cite{Belavin1975}. The winding number is an integer and
reads $Q_{w}=\frac{g^{2}}{32\pi^{2}}\int d^{4}x
G^{a}_{\mu\nu}{\tilde{G}}_{a}^{\mu\nu}$. Here $g$ is the QCD
coupling constant. The gluon field tensor and its dual are
$G^{a}_{\mu\nu}$ and
${\tilde{G}}_{a}^{\mu\nu}=\frac{1}{2}\epsilon^{\mu\nu\rho\sigma}G^{a}_{\rho\sigma}$.
The nontrivial gauge field configurations with non-zero winding
number $Q_{w}$ can result in non-conservation of the axial currents
due to the axial anomaly~\cite{Adler1969}, i.e.,
\begin{equation}
\partial^{\mu}j^{5}_{\mu}=2\sum_{f}m_{f}\langle\bar{\psi}_{f}i\gamma_{5}\psi_{f}\rangle_{A}
-\frac{N_{f}g^{2}}{16\pi^{2}}G^{a}_{\mu\nu}{\tilde{G}}_{a}^{\mu\nu},\label{anomaly}
\end{equation}
where $\psi_{f}$ is a quark field, $m_{f}$ is the current mass of
the quark, and $N_{f}$ is the number of quark flavors.
$j^{5}_{\mu}=\sum_{f}\langle\bar{\psi}_{f}\gamma_{\mu}\gamma_{5}\psi_{f}\rangle_{A}$
denotes the axial current density in the background of a gauge field
configuration $A^{a}_{\mu}$. Integrating the two sides of
Eq.~\eqref{anomaly} over three dimension space and one dimension
time and assuming the number of right-handed and left-handed
fermions is equal initially at $t=-\infty$, we obtain
\begin{equation}
(N_{R}-N_{L})_{t=\infty}=-2N_{f}Q_{w}.\label{NRNL}
\end{equation}
From Eq.~\eqref{NRNL} one can clearly find that through the
interactions between quarks and nontrivial gluon configuration with
non-zero $Q_{w}$, the right-handed quarks are converted into
left-handed quarks, vice versa depending on the sign of the winding
number. Then there will be an asymmetry between the number of right-
and left-handed quarks. When a magnetic field is added, an electric
current is induced along the magnetic field and positive charges are
separated from negative charge, which is called the ``chiral
magnetic
effect''~\cite{Fukushima2008,Kharzeev2009,Kharzeev2010,Fukushima2009,Abramczyk2009,Asakawa2010}.

Now that the chiral magnetic effect can be observed through the
measurements of azimuthal charged-particle correlations in the
relativistic heavy-ion collisions, a natural question arises, i.e.
whether can we detect the properties of the QCD phase transitions,
especially the chiral phase transition through the observations of
the chiral magnetic effect? To answer this question, we have to
study how the chiral magnetic effect or the charge separation effect
is influenced by the chiral phase transition. This is our central
subject in this work.

This work is the extension of our former work~\cite{Fu2010} and
presents many details. The paper is organized as follows. In Sec. II
we simply introduce the thermodynamics of the 2+1 flavor
Polyakov--Nambu--Jona-Lasinio (PNJL) model. In Sec. III we will
calculate the chiral electric current density along the direction of
the magnetic field and study its dependence on the temperature
during the QCD phase transitions. In Sec. IV we will calculate the
electric difference between on each side of the reaction plane and
the azimuthal charged-particle correlations in heavy ion collisions.
In Sec. V we present our summary and conclusions.

\section{Thermodynamics of 2+1 flavor PNJL model}
\label{sec2}

In this work, we will study the chiral magnetic effect and the QCD
phase transitions in the 2+1 flavor Polyakov--Nambu--Jona-Lasinio
model. The validity of the PNJL model has been confirmed in a series
of works by confronting the PNJL results with the lattice QCD
data~\cite{Ratti2006,Roner2007,Ghosh2006,Fu2008,Fu2009}. The PNJL
model not only has the chiral symmetry and the dynamical breaking
mechanism of this symmetry, which are same as the conventional
Nambu--Jona-Lasinio model, but also include the effect of color
confinement through the Polyakov loop. Therefore, the PNJL model is
very appropriate to describe the QCD phase transitions at finite
temperature and/or density.

In this work, we employ the 2+1 flavor Polyakov-loop improved NJL
model which has been discussed in details in our previous work
~\cite{Fu2008}, and the Lagrangian density for the 2+1 flavor PNJL
model is given as
\begin{eqnarray}
\mathcal{L}_{\mathrm{PNJL}}&=&\bar{\psi}(i\gamma_{\mu}D^{\mu}-\hat{m}_{0})\psi
 +G\sum_{a=0}^{8}\Big[(\bar{\psi}\tau_{a}\psi)^{2}\nonumber \\
 &&+(\bar{\psi}i\gamma_{5}\tau_{a}\psi)^{2}\Big]
-K\Big[\textrm{det}_{f}\big(\bar{\psi}(1+\gamma_{5})\psi\big)\nonumber \\
 &&+\textrm{det}_{f}\big(\bar{\psi}(1-\gamma_{5})\psi\big)\Big]
 -\mathcal{U}(\Phi,\Phi^{*} \, ,T),\label{lagragian}
\end{eqnarray}
where $\psi=(\psi_{u},\psi_{d},\psi_{s})^{T}$ is the three-flavor
quark field,
\begin{equation}
D^{\mu}=\partial^{\mu}-iA^{\mu}\quad\textrm{with}\quad
A^{\mu}=\delta^{\mu}_{0}A^{0}\quad\textrm{,}\quad
A^{0}=g\mathcal{A}^{0}_{a}\frac{\lambda_{a}}{2}=-iA_4.
\end{equation}
$\lambda_{a}$ are the Gell-Mann matrices in color space and the
gauge coupling $g$ is combined with the SU(3) gauge field
$\mathcal{A}^{\mu}_{a}(x)$ to define $A^{\mu}(x)$ for convenience.
$\hat{m}_{0}=\textrm{diag}(m_{0}^{u},m_{0}^{d},m_{0}^{s})$ is the
three-flavor current quark mass matrix. Throughout this work, we
take $m_{0}^{u}=m_{0}^{d}\equiv m_{0}^{l}$, while keep $m_{0}^{s}$
being larger than $m_{0}^{l}$, which breaks the $SU(3)_f$ symmetry.
In the above PNJL Lagrangian,
$\mathcal{U}\left(\Phi,\Phi^{*},T\right)$ is the Polyakov-loop
effective potential, which is expressed in terms of the traced
Polyakov-loop $\Phi=(\mathrm{Tr}_{c}L)/N_{c}$ and its conjugate
$\Phi^{*}=(\mathrm{Tr}_{c}L^{\dag})/N_{c}$ with the Polyakov-loop
$L$ being a matrix in color space given explicitly by
\begin{equation}
L\left(\vec{x}\right)=\mathcal{P}\exp\left[i\int_{0}^{\beta}d\tau\,
A_{4}\left(\vec{x},\tau\right)\right] =\exp\left[i \beta A_{4}
\right]\, ,
\end{equation}
with  $\beta=1/T$ being the inverse of temperature and
$A_{4}=iA^{0}$.

In our work, we use the Polyakov-loop effective potential which is a
polynomial in $\Phi$ and $\Phi^{*}$~\cite{Ratti2006}, given by

\begin{equation}
\frac{\mathcal{U}\left(\Phi,\Phi^{*},T\right)}{T^{4}} =
-\frac{b_{2}(T)}{2}\Phi^{*}\Phi -\frac{b_{3}}{6}
(\Phi^{3}+{\Phi^{*}}^{3})+\frac{b_{4}}{4}(\Phi^{*}\Phi)^{2} \, ,
\end{equation}
with
\begin{equation}
b_{2}(T)=a_{0}+a_{1}\left(\frac{T_{0}}{T}\right)+a_{2}
{\left(\frac{T_{0}}{T}\right)}^{2}
+a_{3}{\left(\frac{T_{0}}{T}\right)}^{3}.
\end{equation}
Parameters in the effective potential are fitted to reproduce the
thermodynamical behavior of the pure-gauge QCD obtained from the
lattice simulations. Their values are $a_{0}=6.75$, $a_{1}=-1.95$,
$a_{2}=2.625$, $a_{3}=-7.44$, $b_{3}=0.75$ and $b_{4}=7.5$. The
parameter $T_{0}$ is the critical temperature for the deconfinement
phase transition to take place in the pure-gauge QCD and $T_{0}$ is
chosen to be $270\,\mathrm{MeV}$ according to the lattice
calculations. Furthermore, we also need to determine the five
parameters in the quark sector of the model, which are
$m_{0}^{l}=5.5\;\mathrm{MeV}$, $m_{0}^{s}=140.7\;\mathrm{MeV}$,
$G\Lambda^{2}=1.835$, $K\Lambda^{5}=12.36$ and
$\Lambda=602.3\;\mathrm{MeV}$. They are fixed by fitting
$m_{\pi}=135.0\;\mathrm{MeV}$, $m_{K}=497.7\;\mathrm{MeV}$,
$m_{\eta^{\prime}}=957.8\;\mathrm{MeV}$ and
$f_{\pi}=92.4\;\mathrm{MeV}$~\cite{Rehberg1996}.

In the parity-odd domains which are created during relativistic
heavy-ion collisions, the number of left- and right-hand quarks is
different because of the axial anomaly. In this work we introduce
the chiral chemical potential $\mu_{5}$ to study the left-right
asymmetry following the method of Ref.~\cite{Fukushima2008}, where
the chiral chemical potential $\mu_{5}$ is related with the
effective theta angle of the $\theta$-vacuum through
$\mu_{5}=\partial_{0}\theta/2N_{f}$ and $N_{f}$ is the number of
flavor. Consequently, we should add the following term
\begin{equation}
\bar{\psi}\hat{\mu}_{5}\gamma^{0}\gamma^{5}\psi\label{mu5ter}
\end{equation}
to the Lagrangian density in Eq.~(\ref{lagragian}), where
$\hat{\mu}_{5}=\textrm{diag}(\mu_{5}^{u},\mu_{5}^{d},\mu_{5}^{s})$.
Next, we consider the case that a homogenous magnetic field $B$ is
along the direction of the orbital angular momentum of the system
produced in a non-central heavy-ion collision. In the following we
denote this direction with $z$-direction and particle momentum in
this direction with $p_{3}$. We derive the corresponding
thermodynamics for a system with only one kind of a fermion in
detail in Appendix~\ref{appendixA}. In the Appendixes, we have
emphasized that the approach used in Appendix~\ref{appendixA} is
appropriate to describe the chiral magnetic effect, while the
modified Lagrangian approach in Appendix~\ref{appendixB} is
inappropriate. The results in Appendix~\ref{appendixA} can be easily
generalized to the 2+1 flavor PNJL model and in the mean field
approximation, the thermodynamical potential density
($\Omega=-(T/V)\ln Z$) for the 2+1 flavor quark system under a
homogeneous background magnetic field $B$ and with left-right
asymmetry is given by
\begin{eqnarray}
\Omega &=&-N_{c}\sum_{f=u,d,s}\frac{|q_{f}|e
B}{2\pi}\sum_{n=0}^{\infty}\sum_{s=\pm 1}\int\frac{d
p_{3}}{2\pi}\bigg(E_{f}\nonumber \\
&&+\frac{T}{3}\ln\Big\{1+3\Phi^{*}\exp\big[-\big(E_{f}-\mu_{f}-\frac{s|\epsilon_{f}|}{E_{f}}\mu_{5}^{f}\big)/T\big]\nonumber \\
&&+3\Phi
\exp\big[-2\big(E_{f}-\mu_{f}-\frac{s|\epsilon_{f}|}{E_{f}}\mu_{5}^{f}\big)/T\big]\nonumber\\
&&+\exp\big[-3\big(E_{f}-\mu_{f}-\frac{s|\epsilon_{f}|}{E_{f}}\mu_{5}^{f}\big)/T\big]\Big\}+\frac{T}{3}\ln\Big\{1\nonumber\\
&&+3\Phi
\exp\big[-\big(E_{f}+\mu_{f}-\frac{s|\epsilon_{f}|}{E_{f}}\mu_{5}^{f}\big)/T\big]\nonumber\\
&&+3\Phi^{*}\exp\big[-2\big(E_{f}+\mu_{f}-\frac{s|\epsilon_{f}|}{E_{f}}\mu_{5}^{f}\big)/T\big]\nonumber\\
&&+\exp\big[-3\big(E_{f}+\mu_{f}-\frac{s|\epsilon_{f}|}{E_{f}}\mu_{5}^{f}\big)/T\big]\Big\}\bigg)\nonumber\\
&&+2G({\phi_{u}}^{2}
+{\phi_{d}}^{2}+{\phi_{s}}^{2})-4K\phi_{u}\,\phi_{d}\,\phi_{s}\nonumber\\
&&+\mathcal{U}(\Phi,\Phi^{*},T),\label{thermopotential}
\end{eqnarray}
where we have
\begin{equation}
|\epsilon_{f}|=\sqrt{2n|q_{f}|e B+p_{3}^{2}} \label{}
\end{equation}
and
\begin{equation}
E_{f}=\sqrt{2n|q_{f}|e B+p_{3}^{2}+M_{f}^{2}}.\label{Ef}
\end{equation}
Here $q_{i}(i=u,d,s)$ is the electric charge in unit of elementary
charge $e$ for the quark of flavor $i$. The constituent mass $M_{i}$
is
\begin{equation}
M_{i}=m_{0}^{i}-4G\phi_{i}+2K\phi_{j}\,\phi_{k},\label{constituentmass}
\end{equation}
and $\phi_{i}$ is the chiral condensate
$\langle\bar{\psi}\psi\rangle_{i}$. We also include the quark
chemical potential $\mu_{i}$ in Eq.~\eqref{thermopotential}. From
our calculations in Appendix~\ref{appendixA}, one can find that the
momenta of charged particles in the longitudinal direction, i.e.,
the $z$-direction, are not influenced by the background magnetic
field and $p_{3}$ in the expression of the thermodynamical potential
density in Eq.~\eqref{thermopotential} is continuous; while the
momenta in the transverse plane are discretized due to the external
magnetic field. We should emphasize that at the lowest order of the
transverse quantum number $n=0$, i.e., the lowest order Landau
level, the quark spin only has one value in the $z$-direction, which
means that charged particles in the lowest order Landau level are
polarized by the external magnetic field; however particles in
higher levels, i.e., $n>0$, are not polarized. Therefore, the charge
separation effect only comes from quarks in the lowest order Landau
level.

\section{Chiral electric current along the direction of the magnetic field}
\label{sec3}

The chiral electric current along the longitudinal direction, i.e.
the direction of the magnetic field, is an observable which
describes the magnitude of the charge separation effect. Here, we
use the approach in Appendix~\ref{appendixA} to calculate the chiral
electric current density $j_{3}$, whose expression is
\begin{eqnarray}
j_{3}&=&\frac{e}{V}\int d^{3}x\bar{\psi}\hat{q}\gamma^{3}\psi\nonumber\\
&=&N_{c}\sum_{f=u,d,s}\frac{q_{f}^{2}e^{2}B}{4\pi^{2}}\Big[\int_{0}^{\infty}d
p_{3}\frac{p_{3}}{E_{f}}f\big(E_{f}-\mu_{f}-\frac{p_{3}}{E_{f}}\mu_{5}^{f}\big)\nonumber\\
&&-\int_{0}^{\infty}d
p_{3}\frac{p_{3}}{E_{f}}f\big(E_{f}-\mu_{f}+\frac{p_{3}}{E_{f}}\mu_{5}^{f}\big)+\int_{0}^{\infty}d
p_{3}\frac{p_{3}}{E_{f}}\bar{f}\big(E_{f}\nonumber\\
&&+\mu_{f}-\frac{p_{3}}{E_{f}}\mu_{5}^{f}\big)-\int_{0}^{\infty}d
p_{3}\frac{p_{3}}{E_{f}}\bar{f}\big(E_{f}+\mu_{f}+\frac{p_{3}}{E_{f}}\mu_{5}^{f}\big)\Big],\label{j3three}
\end{eqnarray}
where $\hat{q}e=\mathrm{diag}(q_{u}e,q_{d}e,q_{s}e)$ is the electric
charge matrix for three-flavor quarks; $V$ is the volume of the
system and  $E_{f}$ is given by Eq.~\eqref{Ef} with $n=0$. We have
\begin{equation}
f(x)=\frac{\Phi^{*}e^{-x/T}+2\Phi
e^{-2x/T}+e^{-3x/T}}{1+3\Phi^{*}e^{-x/T}+3\Phi
e^{-2x/T}+e^{-3x/T}}\label{fx}
\end{equation}
and
\begin{equation}
\bar{f}(x)=\frac{\Phi e^{-x/T}+2\Phi^{*}
e^{-2x/T}+e^{-3x/T}}{1+3\Phi e^{-x/T}+3\Phi^{*}
e^{-2x/T}+e^{-3x/T}}.\label{barfx}
\end{equation}
In order to study whether the chiral electric current is affected by
the temperature, quark chemical potential, quark constituent mass
and so on, we just pause here, and turn to the simpler system
composed of only one type of fermion with positive charge $e$ and
mass $m$. The chiral electric current corresponding to this system
is given by
\begin{eqnarray}
j_{3} &=&\frac{e^{2} B}{4\pi^{2}}\Big[\int_{0}^{\infty}d
p_{3}\frac{p_{3}}{E}\frac{1}{e^{(E-\mu-\frac{p_{3}}{E}\mu_{5})/T}+1}-\int_{0}^{\infty}d
p_{3}\frac{p_{3}}{E}\frac{1}{e^{(E-\mu+\frac{p_{3}}{E}\mu_{5})/T}+1}\nonumber\\
&&+\int_{0}^{\infty}d
p_{3}\frac{p_{3}}{E}\frac{1}{e^{(E+\mu-\frac{p_{3}}{E}\mu_{5})/T}+1}-\int_{0}^{\infty}d
p_{3}\frac{p_{3}}{E}\frac{1}{e^{(E+\mu+\frac{p_{3}}{E}\mu_{5})/T}+1}\Big].\label{j3One}
\end{eqnarray}
where
\begin{equation}
E=\sqrt{p_{3}^{2}+m^{2}}.\label{Em}
\end{equation}
It is interesting to consider the case that fermions are massless,
i.e., $m=0$ in Eq.~\eqref{Em}, then it can be easily obtained that
\begin{equation}
j_{3}=\frac{e^{2} B}{2\pi^{2}}\mu_{5}.\label{}
\end{equation}
This is the result obtained in the modified Lagrangian approach as
Eq.~\eqref{j3vac} shows. However, the chiral electric current
calculated in our approach is in essence different from that
obtained in the modified Lagrangian approach. From our calculations
above, one can find that the chiral electric current comes from the
finite temperature part of the thermodynamics, i.e., from fermions
and anti-fermions, whereas the electric current in the modified
Lagrangian approach comes from the Dirac Sea, not from fermions and
anti-fermions (for more details see Appendix~\ref{appendixB}).
Furthermore, although $j_{3}$ obtained in our approach only depends
on the magnetic field strength $B$ and the chiral chemical potential
$\mu_{5}$ in the massless case, it is indeed dependent on the
temperature and the chemical potential $\mu$ when the mass of
fermions is nonvanishing. On the contrary, $j_{3}$ in the modified
approach only depends on $B$ and $\mu_{5}$, regardless of whether
the mass of fermions is vanishing, which is due to the unphysical
ultraviolet momentum in the Dirac Sea.

\begin{figure}[!htb]
\includegraphics[scale=1.2]{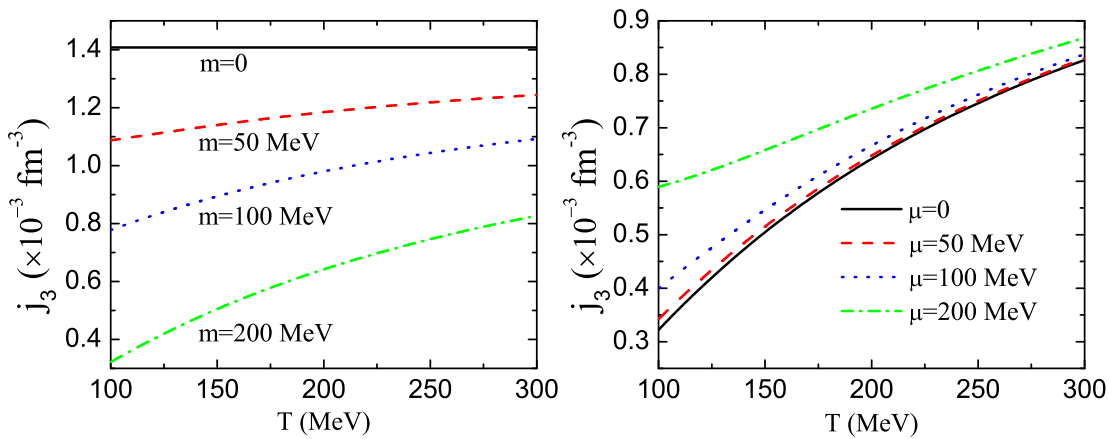}
\caption{(color online). Left panel: chiral electric current density
$j_{3}$ in Eq.~\eqref{j3One} as function of the temperature with
several values of the fermion mass $m$, with
$eB=10^{4}\,\mathrm{MeV}^{2}$, $\mu=0$, and
$\mu_{5}=250\,\mathrm{MeV}$. Right panel: $j_{3}$ as function of the
temperature with several values of the chemical potential $\mu$ and
with $eB=10^{4}\,\mathrm{MeV}^{2}$, $\mu_{5}=250\,\mathrm{MeV}$, and
$m=200\,\mathrm{MeV}$.}\label{f1}
\end{figure}

In order to verify our argument, we calculate the chiral electric
current density in Eq.~\eqref{j3One} for the one fermion system
numerically, and the results are plotted in Fig.~\ref{f1}. From the
figure one can easily find that, although the chiral electric
current density is independent of the temperature when the fermion
is massless as the black solid line in the left panel shows, it is
indeed dependent of the temperature and the chemical potential in
the case that the mass of fermion is nonvanishing. The effect of the
mass of the electric current carrier, i.e., the fermion, is to
decrease $j_{3}$, since the velocity of the fermion becomes smaller
when the mass of the fermion is increased with a fixed energy. On
the contrary, the chiral electric current density increases with the
temperature and the chemical potential.

\begin{figure}[!htb]
\includegraphics[scale=0.8]{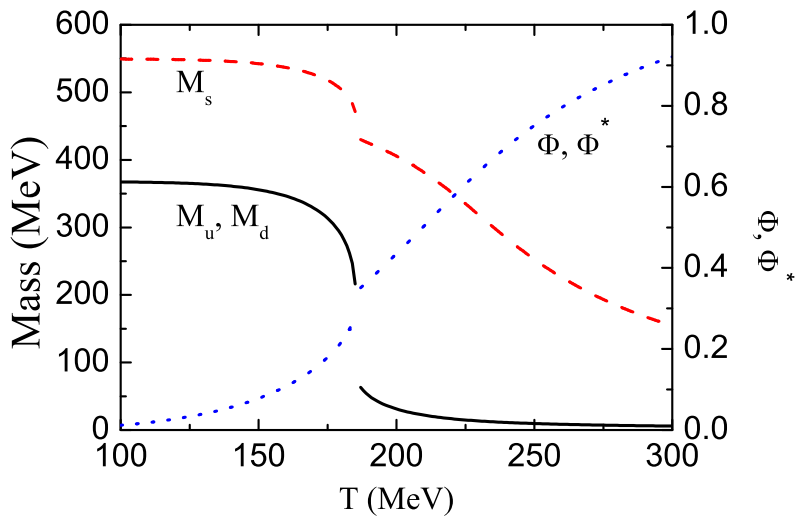}
\caption{(color online). Constituent masses of $u$, $d$ quarks and
$s$ quarks, the Polyakov-loop $\Phi$ and its conjugate $\Phi^{*}$ as
functions of the temperature with $\mu_{i}=0$ ($i=u,d,s$) and
$\mu_{5}=250\,\mathrm{MeV}$
($\mu_{5}\equiv\mu_{5}^{u}=\mu_{5}^{d}=\mu_{5}^{s}$) in the PNJL
model with parameters given in the Sec.\ref{sec2}.}\label{f2}
\end{figure}

Next, we turn our attentions to the chiral electric current produced
in the three-flavor quark system under an external magnetic field,
whose expression is given in Eq.~\eqref{j3three}. From our above
experience that the chiral electric current density would be
affected by the particle mass and temperature, we first investigate
the dependence of the constituent masses of three-flavor quarks on
the temperature during the QCD phase transitions in the PNJL model.
Minimizing the thermodynamical potential in
Eq.~\eqref{thermopotential} with respect to three-flavor quark
condensates, the Polyakov-loop $\Phi$ and its conjugate $\Phi^{*}$,
we obtain a set of equations of motion. We neglect the influence of
the magnetic field on these equations of motion in our numerical
calculations, since the magnetic field
($eB=10^{2}\sim10^{4}\,\mathrm{MeV}^{2}$ in the non-central
heavy-ion collisions~\cite{Kharzeev2008}) has little impact on these
equations of motion. The calculated results are presented in
Fig.~\ref{f2}. Here we take the chiral chemical potential
$\mu_{5}=250\,\mathrm{MeV}$ for example. In the figure we can find
that the constituent masses of $u$, $d$ quarks and $s$ quarks
decrease with the increase of the temperature, and a first order
chiral phase transition takes place at the critical temperature
$T_{C}=185\,\mathrm{MeV}$. The chiral symmetry is restored above
this critical temperature and the constituent masses of $u$, $d$
quarks decrease to their small current masses. Since the $s$ quark
has relatively larger current quark mass, its constituent mass is
still relatively large when the temperature is larger than $T_{C}$,
but it is also decreased quickly with the increase of the
temperature. In our calculations we also note that when the chiral
chemical potential $\mu_{5}$ is decreased, the value of the critical
temperature $T_{C}$ becomes larger and the first order chiral phase
transition gradually evolves to a continuous crossover. Therefore,
the influence of the chiral chemical potential $\mu_{5}$ on the
chiral phase transition is similar with that of the quark chemical
potential $\mu$. In Fig.~\ref{f2} we also plot the Polyakov-loop
$\Phi$ and its conjugate $\Phi^{*}$ versus temperature. One can find
that $\Phi$ and $\Phi^{*}$ increase from 0 to 1 with the increase of
the temperature, implying that the Z(3) symmetry of the gluon field
is broken and the deconfinement phase transition takes
place~\cite{Ratti2006}. We should emphasize that since the term
related with the chiral chemical potential in Eq.~\eqref{mu5ter}
does not broken the charge conjugation symmetry, we have
$\Phi=\Phi^{*}$ even $\mu_{5}$ is nonvanishing, which is different
from the quark chemical potential.

\begin{figure}[!htb]
\includegraphics[scale=0.8]{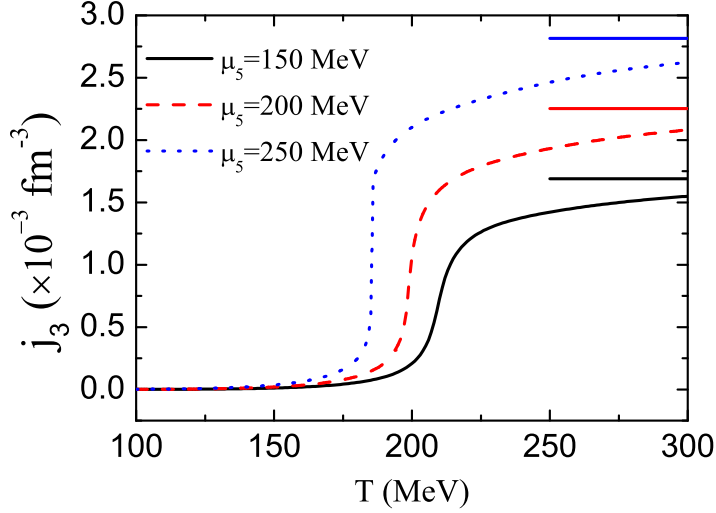}
\caption{(color online). Chiral electric current density along the
direction of an external magnetic field of the three-flavor quark
system $j_{3}$ as function of the temperature in the PNJL model with
$eB=10^{4}\,\mathrm{MeV}^{2}$, $\mu_{i}=0$ ($i=u,d,s$), and several
values of the chiral chemical potential $\mu_{5}$
($\mu_{5}\equiv\mu_{5}^{u}=\mu_{5}^{d}=\mu_{5}^{s}$). The three
horizontal lines denote the values of $j_{3}$ in the high
temperature massless limit, corresponding to $\mu_{5}=$ 150, 200,
and 250 MeVs from bottom to top, respectively.}\label{f3}
\end{figure}

In Fig.~\ref{f3} we show the chiral electric current density of the
three-flavor quark system as function of the temperature in the PNJL
model. Here we take the external magnetic field
$eB=10^{4}\,\mathrm{MeV}^{2}$ for example. First of all, we consider
the high temperature limit. In this limit the masses of quarks can
be neglected and $\Phi=\Phi^{*}=1$. Then the chiral electric current
$j_{3}$ in Eq.~\eqref{j3three} can be easily obtained as
\begin{equation}
j_{3}=N_{c}\Big(\sum_{f=u,d,s}q_{f}^{2}\Big)\frac{e^{2}B\mu_{5}}{2\pi^{2}}=\frac{e^{2}B\mu_{5}}{\pi^{2}}.\label{j3limit}
\end{equation}
We also plot these high temperature limit values of the chiral
electric current density in Fig.~\ref{f3}, i.e. the horizontal lines
from bottom to top corresponding to $\mu_{5}=$ 150, 200, and 250
MeVs, respectively. One can clearly find that when the temperature
is high, the system is in the chiral symmetric and deconfined phase,
and the chiral electric current density approaches its limit value,
i.e. Eq.~\eqref{j3limit}. When the temperature is lowered,
especially when the temperature is below the critical temperature
$T_{C}$, the chiral symmetry is broken and quarks get large
constituent masses, then the chiral electric current density along
the direction of the external magnetic field is quite suppressed and
quickly approaches zero with the decrease of the temperature. This
behavior is independent of the value of the chiral chemical
potential as Fig.~\ref{f3} clearly shows. In Fig.~\ref{f3} we just
take magnetic field strength $eB=10^{4}\,\mathrm{MeV}^{2}$ for
example, and the chiral electric current density is linearly
proportional to the magnetic field strength, since the magnetic
field strength $B$ does not enter into the integrations in the
expression of $j_{3}$, i.e., Eq.~\eqref{j3three}. Therefore, when
the magnetic field strength takes other values, we still have the
fact that when the temperature is below the chiral critical
temperature $T_{C}$, large constituent masses of quarks suppress the
chiral electric current drastically.

\section{Azimuthal charged-particle correlations in heavy-ion collisions}

\begin{figure}[!htb]
\includegraphics[scale=0.6]{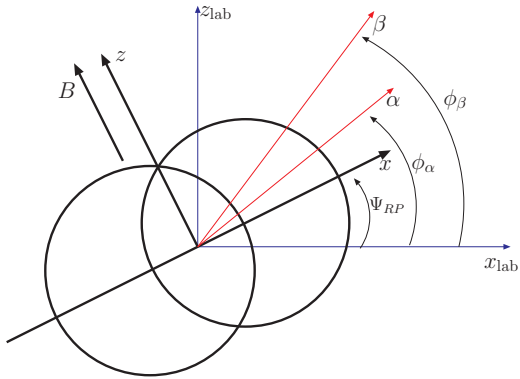}
\caption{(color online). Schematic depiction of the transverse plane
of a non-central heavy ion collision along the beam-axis ($y$-axis)
(see also Ref.~\cite{Abelev2009a}). The plane of $z=0$ is the
reaction plane.}\label{f4}
\end{figure}

In this section we will try to relate our calculations with
experimental observations and investigate how the QCD phase
transitions influence on the signals of the chiral magnetic effect.
In the experiments of heavy ion collisions, the azimuthal
charged-particle correlations, i.e., $\langle
\cos(\phi_{\alpha}+\phi_{\beta}-2\Psi_{RP})\rangle$, are used to
detect the $\mathcal{P}$-violating
effect~\cite{Voloshin2004,Abelev2009a,Abelev2009b}. Here
$\phi_{\alpha}$ and $\phi_{\beta}$ are the azimuthal angles of the
produced particles, and $\alpha$, $\beta$ represent electric charge
$+$ or $-$; $\Psi_{RP}$ is the azimuthal angle of the reaction
plane. These angles are depicted in Fig.~\ref{f4} and in this figure
the reaction plane is the plane of $z=0$ which is perpendicular to
the direction of the magnetic field.

In order to calculate the azimuthal charged-particle correlations,
we follow the method of Ref.~\cite{Kharzeev2008} to define the
quantity $\Delta_{+}$ ($\Delta_{-}$) which is the positive
(negative) electric charge difference in unit of $e$ ($-e$) between
on each side of the reaction plane, i.e., the $z=0$ plane in our
notations.

\begin{figure}[!htb]
\includegraphics[scale=0.6]{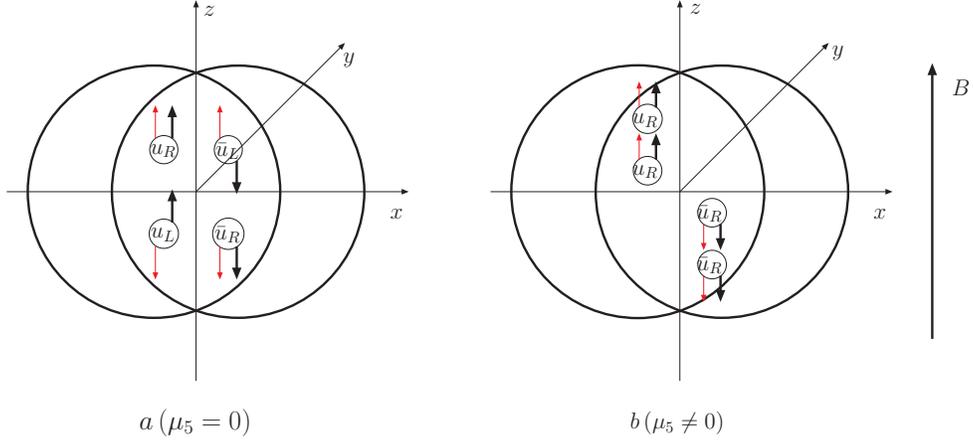}
\caption{(color online). Schematic illustrations of the electric
charge separation and the chiral magnetic effect (see also
Ref.~\cite{Kharzeev2008}). Here we take ``up'' quark $u$ and its
anti-quark $\bar{u}$ for example. The thick black arrows denote the
directions of quark spins and the thin red arrows denote those of
the momentum of quarks. $a.$ Under the external magnetic field,
quarks in the lowest Landau level (not including quarks in high
order Landau levels) are polarized. The spins of $u$ quarks are
parallel to the direction of the magnetic field and those of
$\bar{u}$ anti-parallel to that direction. In the case of
$\mu_{5}=0$, i.e. these is no $\mathcal{P}$-violating effect, the
number of right-handed quarks is equal to that of left-handed quarks
and so the number of quarks above the $z=0$ plane is also equal to
that of quarks below the $z=0$ plane. Therefore, when $\mu_{5}=0$
there is no electric charge separation and chiral magnetic effect.
$b.$ In the case of $\mu_{5}\neq0$, the number of right-handed
quarks is unequal to that of left-handed quarks and the numbers of
quarks on two sides of the $z=0$ plane are different, resulting in
the difference of the electric charges between on each side of the
reaction plane, which is the electric charge separation
effect.}\label{f5}
\end{figure}

In Fig.~\ref{f5} we give an schematic illustration of the chiral
magnetic effect, and detailed discussions are presented in the
caption. From this figure, one can clearly find that it the
$\mathcal{P}$-violating effect, i.e., the nonvanishing chiral
chemical potential $\mu_{5}$, that results in the difference of the
numbers of the right-handed quarks (anti-quarks) and left-handed
quarks (anti-quarks). Then, under an external magnetic field the
number of quarks moving along the direction of the magnetic field
(i.e., the number of the right-hand quarks in Fig.~\ref{f5}) is
different from that of quarks moving against it (i.e., the number of
the left-hand quarks in Fig.~\ref{f5}), and in this way the
phenomenon of the electric charge separation takes place as the
Fig.~\ref{f5} $b$ shows.

Considering the simple system composed of only one type of fermion
(with positive charge $qe$) and anti-fermion once more. From the
Fig.~\ref{f5} and the discussions above, we can easily find that the
difference of the numbers of positive fermions on each side of the
reaction plane is just the difference of the numbers of the
right-handed fermions and the left-handed fermions in the lowest
Landau level, which is just
\begin{eqnarray}
N_{5}|_{n=0}^{+}&=&\Big(\int
d^{3}x{\bar{\psi}}_{R}\gamma^{0}\psi_{R}-\int
d^{3}x{\bar{\psi}}_{L}\gamma^{0}\psi_{L}\Big)\Big|_{n=0}^{+}\nonumber\\
&=&\Big(\int d^{3}x \bar{\psi}\gamma^{0}\gamma^{5}\psi\Big)
\Big|_{n=0}^{+},\label{NN5}
\end{eqnarray}
where
\begin{equation}
\psi_{R}=\frac{1+\gamma^{5}}{2}\psi \quad \mathrm{and}\quad
\psi_{L}=\frac{1-\gamma^{5}}{2}\psi .
\end{equation}
We should emphasize that the subscript $n=0$ in Eq.~\eqref{NN5}
indicates that the difference of the fermion numbers on the two
sides of the reaction plane only comes from fermions in the lowest
Landau level, since only fermions in the lowest Landau level are
polarized, which is proved in Appendix~\ref{appendixA} and
Appendix~\ref{appendixB}. The superscript $+$ in Eq.~\eqref{NN5}
means that only the positive fermions (not the negative
anti-fermions) are included. Therefore, employing Eq.~\eqref{pN5} in
Appendix~\ref{appendixA} we can further express $N_{5}|_{n=0}^{+}$
as
\begin{eqnarray}
N_{5}|_{n=0}^{+}&=&\Big(\int d^{3}x
\bar{\psi}\gamma^{0}\gamma^{5}\psi\Big)
\Big|_{n=0}^{+}\nonumber\\
&=&V\frac{|q|e B}{2\pi}\sum_{s=\pm 1}\int\frac{d
p_{3}}{2\pi}\Big(\frac{s|\epsilon|}{E}\langle{a_{\epsilon}^{s}}^{+}a_{\epsilon}^{s}\rangle\Big)\Big|_{n=0}\nonumber\\
&=&V\frac{|q|e B}{4\pi^{2}}\Big[\int_{0}^{\infty}d
p_{3}\frac{p_{3}}{E}\frac{1}{e^{(E-\mu-\frac{p_{3}}{E}\mu_{5})/T}+1}-\int_{0}^{\infty}d
p_{3}\frac{p_{3}}{E}\frac{1}{e^{(E-\mu+\frac{p_{3}}{E}\mu_{5})/T}+1}\Big],\label{NN5plus}
\end{eqnarray}
where $E$ is given by Eq.~\eqref{Em}. In the same way, we can obtain
the difference of the numbers of negative anti-fermions on each side
of the reaction plane, i.e.,
\begin{eqnarray}
N_{5}|_{n=0}^{-}&=&-\Big(\int d^{3}x
\bar{\psi}\gamma^{0}\gamma^{5}\psi\Big)
\Big|_{n=0}^{-}\nonumber\\
&=&-V\frac{|q|e B}{2\pi}\sum_{s=\pm 1}\int\frac{d
p_{3}}{2\pi}\Big(\frac{s|\epsilon|}{E}\langle{b_{\epsilon}^{s}}^{+}b_{\epsilon}^{s}\rangle\Big)\Big|_{n=0}\nonumber\\
&=&-V\frac{|q|e B}{4\pi^{2}}\Big[\int_{0}^{\infty}d
p_{3}\frac{p_{3}}{E}\frac{1}{e^{(E+\mu-\frac{p_{3}}{E}\mu_{5})/T}+1}-\int_{0}^{\infty}d
p_{3}\frac{p_{3}}{E}\frac{1}{e^{(E+\mu+\frac{p_{3}}{E}\mu_{5})/T}+1}\Big].\label{}
\end{eqnarray}
Until now, we have obtained the difference of numbers of the
positive fermions (negative anti-fermions) between on each side of
the reaction plane, so the electric charge difference can be easily
obtained as $\Delta_{+}=|q|N_{5}|_{n=0}^{+}$ and
$\Delta_{-}=|q|N_{5}|_{n=0}^{-}$. We should emphasize that though
the electric charge differences $\Delta_{+}$ and $\Delta_{-}$ are
the differences of quark electric charges in our picture, these
electric charge differences are conserved through the hadronization
processes (or other processes) and are observed in the heavy ion
collision experiments, because the hadronization processes (or other
processes) are difficult to result in electric charge separations
(for more discussions see Ref.~\cite{Abelev2009a,Abelev2009b}).

The calculations above can be easily extended to the 2+1 flavor
quark system, and for this system we have
\begin{eqnarray}
\Delta_{+}&=&VN_{c}\frac{e
B}{4\pi^{2}}\bigg\{q_{u}^{2}\int_{0}^{\infty}d
p_{3}\frac{p_{3}}{E_{u}}\Big[f\big(E_{u}-\mu_{u}-\frac{p_{3}}{E_{u}}\mu_{5}^{u}\big)-f\big(E_{u}-\mu_{u}+\frac{p_{3}}{E_{u}}\mu_{5}^{u}\big)\Big]\nonumber\\
&&+q_{d}^{2}\int_{0}^{\infty}d
p_{3}\frac{p_{3}}{E_{d}}\Big[\bar{f}\big(E_{d}+\mu_{d}-\frac{p_{3}}{E_{d}}\mu_{5}^{d}\big)-\bar{f}\big(E_{d}+\mu_{d}+\frac{p_{3}}{E_{d}}\mu_{5}^{d}\big)\Big]\nonumber\\&&
+q_{s}^{2}\int_{0}^{\infty}d
p_{3}\frac{p_{3}}{E_{s}}\Big[\bar{f}\big(E_{s}+\mu_{s}-\frac{p_{3}}{E_{s}}\mu_{5}^{s}\big)-\bar{f}\big(E_{s}+\mu_{s}+\frac{p_{3}}{E_{s}}\mu_{5}^{s}\big)\Big]\bigg\},\label{deltaplus}
\end{eqnarray}
and
\begin{eqnarray}
\Delta_{-}&=&-VN_{c}\frac{e
B}{4\pi^{2}}\bigg\{q_{u}^{2}\int_{0}^{\infty}d
p_{3}\frac{p_{3}}{E_{u}}\Big[\bar{f}\big(E_{u}+\mu_{u}-\frac{p_{3}}{E_{u}}\mu_{5}^{u}\big)-\bar{f}\big(E_{u}+\mu_{u}+\frac{p_{3}}{E_{u}}\mu_{5}^{u}\big)\Big]\nonumber\\
&&+q_{d}^{2}\int_{0}^{\infty}d
p_{3}\frac{p_{3}}{E_{d}}\Big[f\big(E_{d}-\mu_{d}-\frac{p_{3}}{E_{d}}\mu_{5}^{d}\big)-f\big(E_{d}-\mu_{d}+\frac{p_{3}}{E_{d}}\mu_{5}^{d}\big)\Big]\nonumber\\&&
+q_{s}^{2}\int_{0}^{\infty}d
p_{3}\frac{p_{3}}{E_{s}}\Big[f\big(E_{s}-\mu_{s}-\frac{p_{3}}{E_{s}}\mu_{5}^{s}\big)-f\big(E_{s}-\mu_{s}+\frac{p_{3}}{E_{s}}\mu_{5}^{s}\big)\Big]\bigg\},\label{deltaminus}
\end{eqnarray}
where $E_{f}$ is given by Eq.~\eqref{Ef} with $n=0$, and the
distribution functions $f(x)$ and $\bar{f}(x)$ for quarks and
anti-quarks respectively, are given by
Eqs.~\eqref{fx}~\eqref{barfx}. One could find that when the quark
chemical potentials are vanishing, i.e., $\mu_{i}=0$ ($i=u,d,s$), we
have $\Delta_{+}=-\Delta_{-}$. In the high temperature limit, the
masses of quarks can be neglected and when the quark chemical
potentials are vanishing, Eq.~\eqref{deltaplus} and
Eq.~\eqref{deltaminus} can be calculated analytically. The results
are
\begin{equation}
\Delta_{+}=-\Delta_{-}=VN_{c}\Big(\sum_{f=u,d,s}q_{f}^{2}\Big)\frac{eB\mu_{5}}{4\pi^{2}}=V\frac{eB\mu_{5}}{2\pi^{2}}.\label{Deltalimit}
\end{equation}

\begin{figure}[!htb]
\includegraphics[scale=0.8]{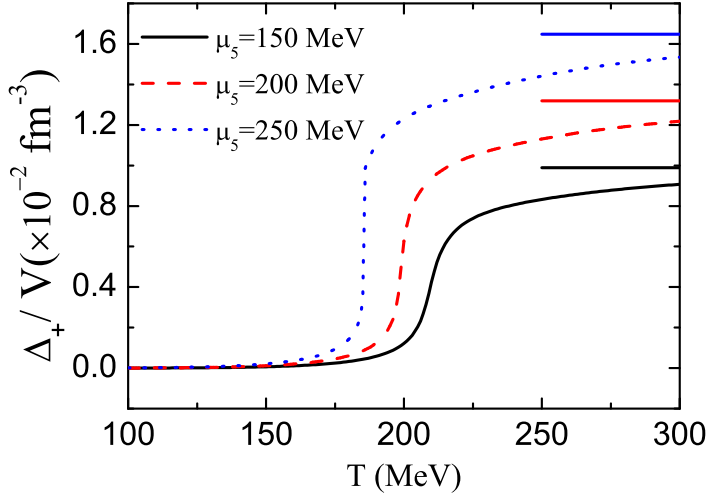}
\caption{(color online). Ratio of the positive electric charge
difference between on each side of the reaction plane and the volume
of the system, i.e.,  $\Delta_{+}/V$, as function of the temperature
in the PNJL model with $eB=10^{4}\,\mathrm{MeV}^{2}$, $\mu_{i}=0$
($i=u,d,s$), and several values of the chiral chemical potential
$\mu_{5}$. The three horizontal lines denote the values of
$\Delta_{+}/V$ in the high temperature massless limit, corresponding
to $\mu_{5}=$ 150, 200, and 250 MeVs from bottom to top,
respectively.}\label{f6}
\end{figure}

In Fig.~\ref{f6} we show $\Delta_{+}/V$, where $V$ is the volume of
the system, as function of the temperature in the PNJL model with
$eB=10^{4}\,\mathrm{MeV}^{2}$, $\mu_{i}=0$, and several values of
the chiral chemical potential $\mu_{5}$. In fact, the dependence of
$\Delta_{+}/V$ on the temperature is similar with that of the chiral
electric current density. When the temperature is above the critical
temperature of the chiral phase transition, $\Delta_{+}/V$
approaches its high temperature limit value given in
Eq.~\eqref{Deltalimit}, which is shown in Fig.~\ref{f6} by the
horizontal lines. However, once the temperature is decreased to that
below the critical temperature, chiral symmetry is broken and quarks
get large constituent masses, which results in that the electric
charge difference between on each side of the reaction plane is
suppressed drastically.

In order to determine the azimuthal charged-particle correlations in
heavy ion collisions, we need to calculate the quantity $N_{+}$
($N_{-}$) which is the total positive (negative) electric charge
number in unit of $e$ ($-e$) on both sides of the reaction plane.
Considering the simple system composed of only one type of fermion
(with positive charge $qe$) and anti-fermion, we can easily find
that the total positive (negative) electric charge is the sum of the
positive (negative) electric charge of the right-handed and
left-handed fermions (anti-fermions), i.e.
\begin{eqnarray}
N_{+}&=&|q|\Big(\int d^{3}x{\bar{\psi}}_{R}\gamma^{0}\psi_{R}+\int
d^{3}x{\bar{\psi}}_{L}\gamma^{0}\psi_{L}\Big)\Big|^{+}\nonumber\\
&=&|q|\Big(\int d^{3}x \bar{\psi}\gamma^{0}\psi\Big)
\Big|^{+}\nonumber\\
&=&V|q|^{2}\frac{e B}{2\pi}\sum_{n=0}^{\infty}\sum_{s=\pm
1}\int\frac{d
p_{3}}{2\pi}\Big(\langle{a_{\epsilon}^{s}}^{+}a_{\epsilon}^{s}\rangle\Big)\nonumber\\
&=&V|q|^{2}\frac{e B}{2\pi}\sum_{n=0}^{\infty}\Big[\int \frac{d
p_{3}}{2\pi}\frac{1}{e^{(E-\mu-\frac{|\epsilon|}{E}\mu_{5})/T}+1}+\int
\frac{d
p_{3}}{2\pi}\frac{1}{e^{(E-\mu+\frac{|\epsilon|}{E}\mu_{5})/T}+1}\Big],\label{NNplus}
\end{eqnarray}
where we have
\begin{equation}
|\epsilon|=\sqrt{2n|q|eB+p_{3}^{2}} \label{}
\end{equation}
and
\begin{equation}
E=\sqrt{2n|q|e B+p_{3}^{2}+m^{2}}.\label{}
\end{equation}
The superscript $+$ on the right hand of the vertical line in
Eq.~\eqref{NNplus} denotes that only fermions with positive charge
are included (not including negative anti-fermions). We should
emphasize that all Landau levels are summed in Eq.~\eqref{NNplus},
which is different from the electric charge difference between on
each side of the reaction plane in Eq.~\eqref{NN5plus}, where only
particles in the lowest Landau level contribute to the charge
asymmetry. In the same way, one can also obtain
\begin{eqnarray}
N_{-}&=&-|q|\Big(\int d^{3}x \bar{\psi}\gamma^{0}\psi\Big)
\Big|^{-}\nonumber\\
&=&V|q|^{2}\frac{e B}{2\pi}\sum_{n=0}^{\infty}\sum_{s=\pm
1}\int\frac{d
p_{3}}{2\pi}\Big(\langle{b_{\epsilon}^{s}}^{+}b_{\epsilon}^{s}\rangle\Big)\nonumber\\
&=&V|q|^{2}\frac{e B}{2\pi}\sum_{n=0}^{\infty}\Big[\int \frac{d
p_{3}}{2\pi}\frac{1}{e^{(E+\mu-\frac{|\epsilon|}{E}\mu_{5})/T}+1}+\int
\frac{d
p_{3}}{2\pi}\frac{1}{e^{(E+\mu+\frac{|\epsilon|}{E}\mu_{5})/T}+1}\Big],\label{NNminus}
\end{eqnarray}

Similarly, for the 2+1 flavor quark system we can obtain
\begin{eqnarray}
N_{+}\!\!&=&\!\!VN_{c}\frac{e B}{2\pi}\sum_{n=0}^{\infty}\sum_{s=\pm
1}\Big[q_{u}^{2}\int\frac{d
p_{3}}{2\pi}f\big(E_{u}-\mu_{u}-\frac{s|\epsilon_{u}|}{E_{u}}\mu_{5}^{u}\big)\nonumber\\
&&+q_{d}^{2}\int\frac{d
p_{3}}{2\pi}\bar{f}\big(E_{d}+\mu_{d}-\frac{s|\epsilon_{d}|}{E_{d}}\mu_{5}^{d}\big)\nonumber\\
&&+q_{s}^{2}\int\frac{d p_{3}}{2\pi}\bar{f}\big(E_{s}+\mu_{s}
-\frac{s|\epsilon_{s}|}{E_{s}}\mu_{5}^{s}\big)\Big].\label{Nplus3}
\end{eqnarray}
and
\begin{eqnarray}
N_{-}\!\!&=&\!\!VN_{c}\frac{e B}{2\pi}\sum_{n=0}^{\infty}\sum_{s=\pm
1}\Big[q_{u}^{2}\int\frac{d
p_{3}}{2\pi}\bar{f}\big(E_{u}+\mu_{u}-\frac{s|\epsilon_{u}|}{E_{u}}\mu_{5}^{u}\big)\nonumber\\
&&+q_{d}^{2}\int\frac{d
p_{3}}{2\pi}f\big(E_{d}-\mu_{d}-\frac{s|\epsilon_{d}|}{E_{d}}\mu_{5}^{d}\big)\nonumber\\
&&+q_{s}^{2}\int\frac{d p_{3}}{2\pi}f\big(E_{s}-\mu_{s}
-\frac{s|\epsilon_{s}|}{E_{s}}\mu_{5}^{s}\big)\Big].\label{Nminus3}
\end{eqnarray}
We should comment that in Eqs.~\eqref{Nplus3}~\eqref{Nminus3} we
have assumed that the total positive (negative) electric charges of
the quarks and anti-quarks in the fireball at early stage are
conserved through the subsequent evolution of the QGP and are
observed by the multiplicities of the produced charged particles in
experiments. Although this is an assumption, it is reasonable.
Because if the centrality is fixed and the collision energy is
increased, on the one hand, the temperature of the QGP at early
stage is increased, which results in that the total positive or
negative electric charges of quarks and anti-quarks increase, on the
other hand, the increase of the collision energy will lead to the
increase of the multiplicities of the produced charged particles.
Therefore, the total positive (negative) electric charges of the
charged particles produced in heavy ion collisions increase with
those of the quarks and anti-quarks.

So far, we can calculate the azimuthal charged particle correlations
$\langle \cos(\phi_{\alpha}+\phi_{\beta}-2\Psi_{RP})\rangle$ in
heavy ion collisions. With the notation
$a_{\alpha\beta}\equiv-\langle
\cos(\phi_{\alpha}+\phi_{\beta}-2\Psi_{RP})\rangle$, it can be shown
that~\cite{Kharzeev2008}
\begin{equation}
a_{++}=\frac{\pi^{2}}{16}\frac{\langle\Delta_{+}^{2}\rangle}{N_{+}^{2}},
\quad
a_{--}=\frac{\pi^{2}}{16}\frac{\langle\Delta_{-}^{2}\rangle}{N_{-}^{2}},\label{a1}
\end{equation}
and
\begin{equation}
a_{+-}=\frac{\pi^{2}}{16}\frac{\langle\Delta_{+}\Delta_{-}\rangle}{N_{+}N_{-}},\label{a2}
\end{equation}
where the azimuthal angle distribution of the charged particles is
assumed to be
\begin{equation}
\frac{d N_{\pm}}{d
\phi}=\frac{1}{2\pi}N_{\pm}+\frac{1}{4}\Delta_{\pm}\sin(\phi-\Psi_{RP}).
\end{equation}
Since we mainly focus on the influence of the QCD phase transitions,
especially the chiral phase transition, on the chiral magnetic
effect in this work, we will neglect the screening suppression
effect due to the final state interactions~\cite{Kharzeev2008} and
make $\mu_{i}=0$ ($i=u,d,s$), then we have $a_{++}=a_{--}=-a_{+-}$.
Therefore, we only study $a_{++}$ in the following.

\begin{figure}[!htb]
\includegraphics[scale=0.8]{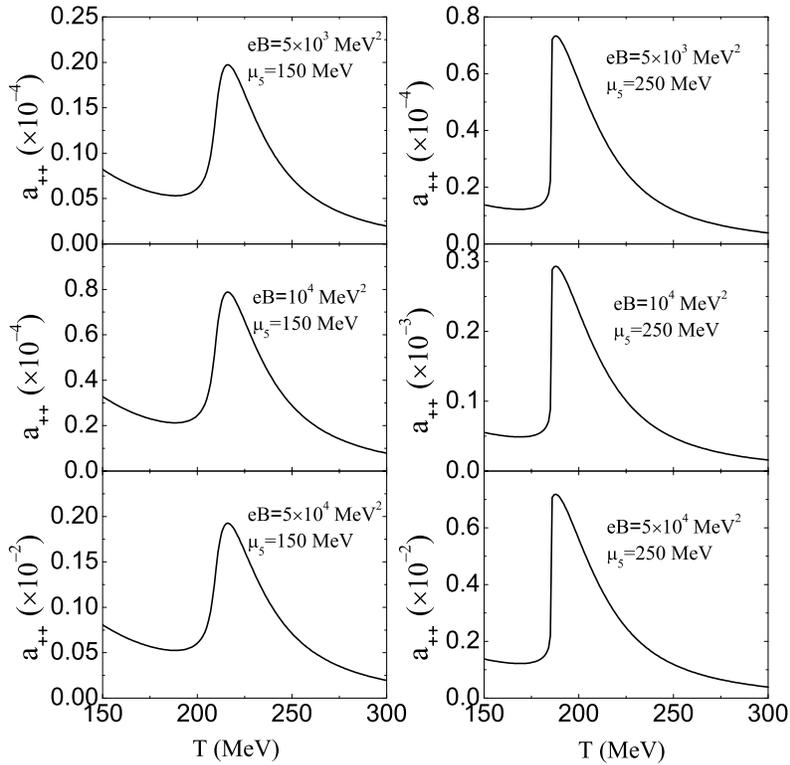}
\caption{Correlation $a_{++}$ as function of the temperature
calculated in the PNJL model with $\mu_{5}=150\,\mathrm{MeV}$ (left
panel) and $\mu_{5}=250\,\mathrm{MeV}$ (right panel). The magnetic
field corresponds to $e B=5\times 10^{3}$, $10^{4}$, and $5\times
10^{4}\,\mathrm{MeV}^{2}$ from top to bottom,
respectively.}\label{f7}
\end{figure}

In Fig.~\ref{f7} we show $a_{++}$ defined in Eq.~(\ref{a1}) as
function of the temperature at several values of the chiral chemical
potential $\mu_{5}$
($\mu_{5}\equiv\mu_{5}^{u}=\mu_{5}^{d}=\mu_{5}^{s}$) and the
magnetic field strength. We find that there is a pronounced cusp in
$a_{++}$ at the critical temperature during the chiral phase
transition (the critical temperature $T_{c}=209\,\mathrm{MeV}$ for
$\mu_{5}=150\,\mathrm{MeV}$ and $T_{c}=185\,\mathrm{MeV}$ for
$\mu_{5}=250\,\mathrm{MeV}$ in the PNJL model). From the
Fig.~\ref{f7} one can also find that although the value of $a_{++}$
is proportional to the square of the magnetic field strength, the
shape of the curve for $a_{++}$ as function of temperature is almost
independent of the magnetic field strength. Furthermore, the cusp at
the critical temperature in the curve becomes much sharper with the
increase of the chiral chemical potential. With the decrease of the
temperature, when the temperature is below $T_{c}$, chiral symmetry
is dynamically broken and quarks obtain large constituent masses. We
should emphasize that it is the large quark mass that results in the
drastic suppression of the chiral electric current density (see
Sec.~\ref{sec3}), the electric charge difference between on each
side of the reaction plane (see Fig.~\ref{f6}), and the azimuthal
charged particle correlations. Furthermore, the chiral magnetic
effect is close related with the axial anomaly~\cite{Kharzeev2008}.
Without axial anomaly there would not be the chiral magnetic effect.
Since the axial anomaly can be suppressed by the mass effect, which
has been discussed in detail in Ref.~\cite{Ma2006}, it is natural to
expect that the chiral magnetic effect can also be suppressed by
large constituent quark masses. Therefore, it is reasonable that the
azimuthal charged particle correlations described by $a_{++}$
($a_{--}$ and $a_{+-}$) defined in Eqs.~(\ref{a1})~(\ref{a2}) are
quite decreased once the temperature is below the critical
temperature. It can been seen from Fig.~\ref{f7} that, when the
temperature is above $T_{c}$, $a_{++}$ decreases with the increase
of the temperature, which is because higher temperature makes it
more difficult to polarize quarks with magnetic field and thus
suppresses the charge separation effect.

What do our calculated results imply in future energy scanning
experiments of heavy ion collisions? With the decrease of the heavy
ion collision energy, the temperature of the QGP produced in the
fireball at early stage is also decreased. Since the magnetic field
produced in non-central collisions decays with
time~\cite{Kharzeev2008}, the observed charge separation mainly
carries the information of the QGP at early stage. Therefore, we
expect that when the collision energy is decreased to a value that
cannot drive the chiral phase transition, the azimuthal charged
particle correlations (especially for the same charge correlations,
because the opposite charge correlations are suppressed by final
state interactions) are quite suppressed. So this property can be
employed to search for where the QCD phase transitions take place.

From the calculations of the azimuthal charged particle correlations
$\langle \cos(\phi_{\alpha}+\phi_{\beta}-2\Psi_{RP})\rangle$ above,
we can find that there are some uncertainties on the total positive
(negative) electric charges of the produced charged particles
$N_{+}$ ($N_{-}$). This little defect motivate us to search for
other better correlators which are not divided by the square of
$N_{+}$ or $N_{-}$. In fact, this kind of correlators has been
proposed by D.E.Kharzeev and his collaborators~\cite{Kharzeev2008}.
In the following we compare these two kinds of correlators briefly.
First of all, we make $(\phi_{\alpha}-\Psi_{RP})\rightarrow
\phi_{\alpha}$ and $(\phi_{\beta}-\Psi_{RP})\rightarrow
\phi_{\beta}$, and then the $\phi_{\alpha}$ and $\phi_{\beta}$ are
the azimuthal angles of produced particles with respective to the
reaction plane as Fig.~\ref{f4} shows. For each collision event, we
follow the definition of the correlators in
Ref.~\cite{Kharzeev2008}, i.e.,
\begin{equation}
f(\phi_{\alpha},\phi_{\beta})=\frac{1}{N_{\alpha}N_{\beta}}
\sum_{i=1}^{N_{\alpha}}\sum_{j=1}^{N_{\beta}}\cos(\phi_{\alpha
i}+\phi_{\beta j}).
\end{equation}
In the same way, here $\alpha,\beta=\pm$ denotes the electric
charge. In order to remove the multiplicity fluctuations the
correlators are averaged over $N_{e}$ similar events. Then one
obtain
\begin{eqnarray}
a_{\alpha\beta}&=&-\langle \cos(\phi_{\alpha}+\phi_{\beta})\rangle\nonumber\\
&=&-\frac{1}{N_{e}}\sum_{n=1}^{N_{e}}f(\phi_{\alpha},\phi_{\beta}),
\end{eqnarray}
where the correlators $a_{\alpha\beta}$ are those calculated in
Fig.~\ref{f7}. Furthermore, correlators which are not divided by
square of the total multiplicity of charged particles are also
proposed by D.E.Kharzeev and his collaborators and are thought to be
very useful, which are
\begin{equation}
b_{\alpha\beta}=-\frac{1}{N_{e}}\sum_{n=1}^{N_{e}}g(\phi_{\alpha},\phi_{\beta}),
\end{equation}
with
\begin{equation}
g(\phi_{\alpha},\phi_{\beta})=\sum_{i=1}^{N_{\alpha}}\sum_{j=1}^{N_{\beta}}\cos(\phi_{\alpha
i}+\phi_{\beta j}).
\end{equation}
Same as the $a_{\alpha\beta}$, it can be found that
\begin{equation}
b_{++}=\frac{\pi^{2}}{16}\langle\Delta_{+}^{2}\rangle, \quad
b_{--}=\frac{\pi^{2}}{16}\langle\Delta_{-}^{2}\rangle,
\end{equation}
and
\begin{equation}
b_{+-}=\frac{\pi^{2}}{16}\langle\Delta_{+}\Delta_{-}\rangle.
\end{equation}
In Fig.~\ref{f6} we have calculated the $\Delta_{+}/V$ as function
of the temperature during the QCD phase transitions, and find that
$\Delta_{+}/V$ is rapidly suppressed and approaches zero when the
temperature is below the chiral critical temperature. Within similar
collision events (similar centrality and atomic number), it can be
expected that the dependence of the volume of the fireball at the
early stage on the collision energy is mild. Therefore, it can be
predicted that with the decrease of the collision energy, the
correlators $b_{\alpha\beta}$, which are the azimuthal charged
particle correlations not divided by the square of the total
multiplicity of charged particles, get a sudden suppression at the
critical temperature of the QCD phase transitions and approaches
zero rapidly.

\section{Summary and Discussions}

In this work, we have studied the influence of the QCD phase
transitions on the chiral magnetic effect. The chiral electric
current density, the electric charge difference between on each side
of the reaction plane, and the azimuthal charged particle
correlations in heavy ion collisions are calculated in the PNJL
model, and their dependence on the temperature are studied. We find
that with the decrease of the temperature, the chiral electric
current density and the electric charge difference between on each
side of the reaction plane are suppressed abruptly at the critical
temperature of the QCD phase transitions and approach zero rapidly,
since below the critical temperature the chiral symmetry is broken
and quarks obtain large constituent mass. It is the large quark mass
that suppresses not only the axial anomaly but also the chiral
magnetic effect. For the azimuthal charged particle correlations, we
study not only the correlators $a_{\alpha\beta}$, which are the
correlators divided by the square of the total multiplicity of
charged particles and are measured in current experiments, but also
another kind of correlators $b_{\alpha\beta}$ which are not divided
by the square of the total multiplicity. We find that both
$a_{\alpha\beta}$ and $b_{\alpha\beta}$ get a sudden suppression at
the critical temperature of the QCD phase transitions. Furthermore,
the correlators $b_{\alpha\beta}$ approaches zero rapidly once the
temperature decreases to values that are below the critical
temperature. Therefore, It indicates that azimuthal charged particle
correlations (both $a_{\alpha\beta}$ and $b_{\alpha\beta}$, in fact
$b_{\alpha\beta}$ is better because the correlators
$b_{\alpha\beta}$ remove the fluctuations of the total multiplicity
of charged particles) can be used as a signal to identify the chiral
phase transition in the energy scan experiment in RHIC.

We should discuss the possibility that using the azimuthal charged
particle correlations ($a_{\alpha\beta}$ and $b_{\alpha\beta}$) to
search for where the QCD phase transitions take place in future
energy scanning experiments in RHIC. In order to simplify the
calculations, we make the magnetic field strength and the chiral
chemical potential fixed across the QCD phase transitions in this
work. In the realistic situations the magnetic field decays with
time and the chiral chemical potential has some distribution.
However, we think that our simplification is reasonable and would
not change our conclusions. The reasons are listed below:
\begin{description}
\item{(1)} In this work we are concentrated on the influence of
the QCD phase transitions, especially the chiral phase transition,
on the chiral magnetic effect embodied by the phenomenon of the
charge separation. Since the chiral phase transition takes place
during a very narrow region of the temperature (or the collision
energy equivalently) as Fig.~\ref{f2} shows, the dependence of the
magnetic field and the chiral chemical potential on the temperature
is limited in this narrow region.
\item{(2)} Indeed the magnetic field decays with time in a
collision event, which has been confirmed in
Ref.~\cite{Kharzeev2008}. However, what influences on our
calculations is the dependence of the magnetic field in the fireball
at early stage on the collision energy (different events with
different collision energy). Why is the magnetic field at the early
stage of the evolution of the fireball? This is because the electric
charge difference between on each side of the reaction plane
$\Delta_{\pm}$ (describing the magnitude of the charge separation)
is proportional to the magnetic field strength as
Eqs.~\eqref{deltaplus}~\eqref{deltaminus} show (we should emphasize
that the total multiplicity of the charged particle $N_{\pm}$ is
almost not affected by the magnetic field with value $eB=10^{2}\sim
10^{4}\,\mathrm{MeV}^{2}$ in the non-central heavy ion collisions).
Therefore, the electric charge difference coming from the early
stage of the fireball evolution is much larger than that from the
late stage, since the magnetic field at early stage is larger than
that at late stage. So we are more concerned about the magnetic
field at early stage. Fortunately, for similar collision events
(similar centrality and atomic number), the dependence of the
magnetic field on the collision energy is very mild (comparing
Fig.A.1. (center of mass energy per nucleon pair being
$\sqrt{s}=62\,\mathrm{GeV}$) with Fig.A.2 (
$\sqrt{s}=200\,\mathrm{GeV}$) in Ref.~\cite{Kharzeev2008}). This is
because what determines the magnetic field is the velocity of the
heavy ion for similar collision events. However, for
$\sqrt{s}=200\,\mathrm{GeV}$ the velocity of the heavy ion is
$v=0.99995c$ where $c$ is the light speed; for
$\sqrt{s}=62\,\mathrm{GeV}$ the velocity of the heavy ion is
$v=0.99948c$. Therefore, although the difference of the two
collision energy is quite large, the difference of their
corresponding velocity of the heavy ion is quite small, which
results in that the difference of the magnetic field is small.
\item{(3)} As for the chiral chemical potential, we should comment
that in the chiral symmetry broken phase, quarks get constituent
mass. It is found that the mass always causes the asymmetry between
the number of right-handed and left-handed fermions to
decay~\cite{Ambjorn1983}, i.e., the chiral chemical potential
decreases with time in the chiral symmetry broken phase. Our
calculations above indicate that when the temperature is decreased
and crosses the critical temperature of the chiral phase transition,
large quark constituent mass suddenly suppresses the electric charge
difference between on each side of the reaction plane. These
calculations are performed with the chiral chemical potential fixed.
If we further consider that the chiral chemical potential is reduced
when the temperature is below the critical temperature, the
suppression is much more significant.
\item{(4)} In this work we perform our calculations with
multi-values for the magnetic field strength and the chiral chemical
potential (see Fig.~\ref{f3}, Fig.~\ref{f6}, and Fig.~\ref{f7}). For
all these values we find the same conclusion that in the chiral
symmetry broken phase, the chiral magnetic effect is quite
suppressed and almost vanishes. Therefore, the chiral magnetic
effect can be used as an order parameter of the QCD phase
transitions.
\item{(5)} We should emphasize that the physical essence underlying
our calculated results is very important. It is the large mass that
suppresses the axial anomaly and the asymmetry between the number of
right-handed and left-handed fermions, which is verified in general
quantum field theory~\cite{Ma2006,Ambjorn1983}. When the mass of the
fermion approaches infinity, there would be of course no difference
between right- and left-handed fermions. Therefore, our calculated
results is consistent with this basic principle and verify the
conjecture proposed by D.E.Kharzeev and his collaborators that the
chiral magnetic effect can be used as an order parameter for the QCD
phase transition~\cite{Kharzeev2008}.
\end{description}

\begin{acknowledgments}
This work was supported by the National Natural Science Foundation
of China under contract Nos. 10425521, 10675007, 10935001, the Major
State Basic Research Development Program under contract Nos.
G2007CB815000. One of the authors (W.J.F.) would also acknowledge
the financial support from China Postdoctoral Science Foundation No.
20090460534.
\end{acknowledgments}

\appendix

\section{Thermodynamics of a fermion system with $\mathcal{P}$ violation and under a background magnetic field}
\setcounter{section}{1} \setcounter{equation}{0} \label{appendixA}

Considering a system composed of only one type of fermion with
positive charge $e$ and mass $m$, and a homogeneous magnetic field
with strength $B$ is along the positive $z$ direction. Assuming the
system is in thermodynamical equilibrium with temperature $T$ and
chemical potential $\mu$. In order to include the effects of
$\mathcal{P}$ and $\mathcal{CP}$ violation, we follow the method of
Ref.~\cite{Fukushima2008} to introduce the chiral chemical potential
$\mu_{5}$. We begin with the partition function of the system as
\begin{equation}
Z=\mathrm{Tr}e^{-\beta(\hat{H}-\mu
\hat{N}-\mu_{5}\hat{N_{5}})},\label{partition_function}
\end{equation}
where $\beta=1/T$ and quantities with hat are operators, and the
Hamiltonian $\hat{H}$ is
\begin{equation}
\hat{H}=\int d^{3}x \mathcal{H}=\int d^{3}x
\bar{\psi}(-i\gamma^{i}\partial_{i}-e\gamma^{i}A^{i}+m)\psi,\label{Hamiltonian}
\end{equation}
here $i=1,2,3$. The Hamiltonian density $\mathcal{H}$ above can be
obtained from the lagrangian density given by
\begin{equation}
\mathcal{L}=\bar{\psi}(i\gamma^{\mu}D_{\mu}-m)\psi,\label{}
\end{equation}
where $D_{\mu}=\partial_{\mu}+ieA_{\mu}$ and $\mu=0,1,2,3$. In the
Hamiltonian Eq.~\eqref{Hamiltonian}, we have used the fact that
since we only consider the case with homogenous magnetic field along
the positive $z$ direction, we can assume $A_{1}=-(1/2)Bx_{2}$ and
$A_{2}=(1/2)Bx_{1}$. $\hat{N}$ and $\hat{N_{5}}$ in the partition
function in Eq.~\eqref{partition_function} are
\begin{eqnarray}
\hat{N}&=&\int d^{3}x\bar{\psi}\gamma^{0}\psi,  \\
\hat{N_{5}}&=&\int d^{3}x\bar{\psi}\gamma^{0}\gamma^{5}\psi,
\end{eqnarray}
respectively.

First of all, we should solve the Dirac equation with a magnetic
field, i.e.
\begin{equation}
i\partial_{0}\psi=\Big[(-i\partial_{i}-eA^{i})\gamma^{0}\gamma^{i}+\gamma^{0}m\Big]\psi.\label{Dirac_equation}
\end{equation}
In the following, we employ the notations in
Ref.~\cite{Metlitski2005} and use the chiral representation of the
$\gamma$ matrices, i.e.
\begin{equation}
\gamma^{0}=\Bigg(\begin{array}{cc}0&1\\
1&0\end{array}\Bigg),\quad\gamma^{i}=\Bigg(\begin{array}{cc}0&\sigma^{i}\\
-\sigma^{i}&0\end{array}\Bigg),\quad\gamma^{5}=\Bigg(\begin{array}{cc}-1&0\\
0&1\end{array}\Bigg).\label{}
\end{equation}
We express the four-component spinor as two two-component
left-handed and right handed Weyl spinors, i.e.
\begin{equation}
\psi=\Bigg(\begin{array}{c}\psi_{L}\\
\psi_{R}\end{array}\Bigg).\label{}
\end{equation}
Then, for the positive energy solution, the Dirac
equation~\eqref{Dirac_equation} can be expressed as
\begin{equation}
\Bigg(\begin{array}{cc}-(-i\partial_{i}-eA^{i})\sigma^{i}& m\\
m & (-i\partial_{i}-eA^{i})\sigma^{i}\end{array}\Bigg)\Bigg(\begin{array}{c}\psi_{L}\\
\psi_{R}\end{array}\Bigg)=E\Bigg(\begin{array}{c}\psi_{L}\\
\psi_{R}\end{array}\Bigg).\label{}
\end{equation}
We set $H_{+}\equiv (-i\partial_{i}-eA^{i})\sigma^{i}$, then if we
find an appropriate right handed spinor $\psi_{R}$, which is an
eigenfunction of the $H_{+}$ with eigenvalue $\epsilon$, i.e.
\begin{equation}
H_{+}\psi_{R}=\epsilon\psi_{R},\label{}
\end{equation}
we have $E^{2}=\epsilon^{2}+m^{2}$ and
\begin{equation}
\psi_{L}=\frac{(E-\epsilon)}{m}\psi_{R}.\label{}
\end{equation}
We set $\psi_{R}=\sqrt{E+\epsilon}\xi^{s}$, therefore, we have
$\psi_{L}=\sqrt{E-\epsilon}\xi^{s}$ and
$H_{+}\xi^{s}=\epsilon\xi^{s}$.

For the negative energy solution of the Dirac
equation~\eqref{Dirac_equation}, we have
\begin{equation}
\Bigg(\begin{array}{cc}H_{-}& m\\
m & -H_{-}\end{array}\Bigg)\Bigg(\begin{array}{c}\psi_{L}\\
\psi_{R}\end{array}\Bigg)=-E\Bigg(\begin{array}{c}\psi_{L}\\
\psi_{R}\end{array}\Bigg).\label{Diracmin}
\end{equation}
where we set $H_{-}\equiv-H_{+}=(i\partial_{i}+eA^{i})\sigma^{i}$.
In the same way, if we find a $\xi^{-(s)}$ satisfying
$H_{-}\xi^{-(s)}=\epsilon\xi^{-(s)}$, then we can obtain
\begin{equation}
\psi=\Bigg(\begin{array}{c}\sqrt{E-\epsilon}\xi^{-(s)}\\
-\sqrt{E+\epsilon}\xi^{-(s)}\end{array}\Bigg).\label{}
\end{equation}

In the following, we will solve the eigenvalue equation
\begin{equation}
H_{+}\xi^{s}=\epsilon\xi^{s},\label{Hplus}
\end{equation}
and here,
\begin{eqnarray}
H_{+}&=&(-i\partial_{i}-eA^{i})\sigma^{i} \nonumber \\
&=&-i\partial_{3}\sigma^{3}+(-i\partial_{a}-eA^{a})\sigma^{a}\nonumber\\
&=&p^{3}\sigma^{3}+H_{\bot},
\end{eqnarray}
where $a=1,2$ and in the last line we have used the fact that
$A^{i}$ is independent of $x_{3}$ and $A^{3}=0$, so the eigenstates
in the $x_{3}$ direction are free continuum of momentum. We should
note that since $\{\sigma^{3}, H_{\bot}\}=0$, if there is a
eigenstate $|\lambda\rangle$ of $H_{\bot}$ with eigenvalue $\lambda
>0$, there must be another eigenstate $\sigma^{3}|\lambda\rangle$ of
$H_{\bot}$ with eigenvalue $-\lambda <0$. In the representation of
$|\lambda\rangle$ and $\sigma^{3}|\lambda\rangle$, Eq.~\eqref{Hplus}
can be expressed as
\begin{equation}
\Bigg(\begin{array}{cc}\lambda & p^{3}\\
p^{3} & -\lambda\end{array}\Bigg)\Bigg(\begin{array}{c}c_{1}\\
c_{2}\end{array}\Bigg)=\epsilon\Bigg(\begin{array}{c}c_{1}\\
c_{2}\end{array}\Bigg),\label{matrix}
\end{equation}
where $\xi^{s}=(c_{1},c_{2})^{T}$. Eq.~\eqref{matrix} has two
solutions which are
\begin{equation}
\xi^{1}=\frac{1}{\sqrt{2|\epsilon|}}\Bigg(\begin{array}{c}\mathrm{sgn}(p_{3})\sqrt{|\epsilon|+\lambda}\\
\sqrt{|\epsilon|-\lambda}\end{array}\Bigg),\quad\mathrm{with}\quad
\epsilon=|\epsilon|,\label{xi1}
\end{equation}
and
\begin{equation}
\xi^{-1}=\frac{1}{\sqrt{2|\epsilon|}}\Bigg(\begin{array}{c}-\mathrm{sgn}(p_{3})\sqrt{|\epsilon|-\lambda}\\
\sqrt{|\epsilon|+\lambda}\end{array}\Bigg),\quad\mathrm{with}\quad
\epsilon=-|\epsilon|,\label{xi2}
\end{equation}
where $|\epsilon|=\sqrt{\lambda^{2}+p_{3}^{2}}$ and $\xi^{s}$ has
been normalized. Here we have use $p_{3}$ to stand for $p^{3}$
without confusion. Equations~\eqref{xi1}~\eqref{xi2} can also be
unified to express as
\begin{equation}
\xi^{s}=\frac{1}{\sqrt{2|\epsilon|}}\Bigg(\begin{array}{c}(s)\mathrm{sgn}(p_{3})\sqrt{|\epsilon|+s\lambda}\\
\sqrt{|\epsilon|-s\lambda}\end{array}\Bigg),\quad\mathrm{with}\quad
\epsilon=s|\epsilon|.\label{xi}
\end{equation}
In the same way, we can solve $H_{-}\xi^{-(s)}=\epsilon\xi^{-(s)}$
for the anti-particle, i.e.,
\begin{equation}
\xi^{-(s)}=\frac{1}{\sqrt{2|\epsilon|}}\Bigg(\begin{array}{c}(-s)\mathrm{sgn}(p_{3})\sqrt{|\epsilon|-s\lambda}\\
\sqrt{|\epsilon|+s\lambda}\end{array}\Bigg),\quad\mathrm{with}\quad
\epsilon=-s|\epsilon|.\label{}
\end{equation}

Next, we turn to the eigen-equation of the transverse momentum
$H_{\perp}|\lambda\rangle=\lambda|\lambda\rangle$. It is obvious
that we also have
\begin{equation}
H_{\perp}^{2}|\lambda\rangle=\lambda^{2}|\lambda\rangle.\label{H2}
\end{equation}
$H_{\perp}^{2}$ can be directly calculated as
\begin{eqnarray}
H_{\perp}^{2}&=&[(-i\partial_{a}-eA^{a})\sigma^{a}]^{2}\nonumber\\
&=&\big(-i\frac{\partial}{\partial
x_{1}}\big)^{2}+\big(-i\frac{\partial}{\partial
x_{2}}\big)^{2}+\frac{1}{4}e^{2}B^{2}(x_{1}^{2}+x_{2}^{2})\nonumber\\
&&-eB\big[x_{1}\big(-i\frac{\partial}{\partial
x_{2}}\big)-x_{2}\big(-i\frac{\partial}{\partial
x_{1}}\big)\big]-eB\sigma^{3}\label{transverse}.
\end{eqnarray}
The physical meanings of Eq.~\eqref{transverse} are very clear. The
second line of Eq.~\eqref{transverse} indicates that the dynamics of
particles in the transverse plane, which is perpendicular to the
magnetic field, is the two dimensional homogeneous harmonic
oscillation. The last line of Eq.~\eqref{transverse} includes
contributions from the orbital and spin angular momentum in the $z$
direction. Rescaling $x_{1}\rightarrow(eB/2)^{1/2}x_{1}$,
$x_{2}\rightarrow(eB/2)^{1/2}x_{2}$, and$H_{\perp}^{2}\rightarrow
H_{\perp}^{2}/eB$, we find
\begin{equation}
H_{\perp}^{2}=\frac{1}{2}[(p_{1}^{2}+x_{1}^{2})+(p_{2}^{2}+x_{2}^{2})]-(l_{3}+\sigma^{3}),\label{}
\end{equation}
where $p_{a}=-i\partial/\partial x_{i}$ ($i=1,2$) and
$l_{3}=x_{1}p_{2}-x_{2}p_{1}$. In the following, we use the
algebraic method to solve the problem of eigenvalue. Introducing
annihilation and creation operators
\begin{eqnarray}
a_{i}&=&\frac{1}{\sqrt{2}}(x_{i}+i p_{i}),\nonumber\\
a_{i}^{+}&=&\frac{1}{\sqrt{2}}(x_{i}-i p_{i}).
\end{eqnarray}
It can be easily verified that
\begin{equation}
[a_{i},a_{j}]=0\quad\mathrm{and}\quad
[a_{i},a_{j}^{+}]=\delta_{ij},\label{}
\end{equation}
thus these operators are annihilation and creation operators of
boson. Employing these operators we obtain
\begin{equation}
H_{\perp}^{2}=a_{1}^{+}a_{1}+a_{2}^{+}a_{2}+1-(l_{3}+\sigma^{3}),\label{}
\end{equation}
with
\begin{equation}
l_{3}=i(-a_{1}^{+}a_{2}+a_{2}^{+}a_{1}).\label{}
\end{equation}
To diagonalize the orbital angular momentum $l_{3}$, we introduce
two another pairs of annihilation and creation operators, and the
annihilation operators are
\begin{eqnarray}
a_{+}&=&\frac{1}{\sqrt{2}}(a_{1}-i a_{2}),\nonumber\\
a_{-}&=&\frac{1}{\sqrt{2}}(a_{1}+i a_{2}).
\end{eqnarray}
In the same way, it can be verified that they are also bosonic
operators. Consequently, we can obtain
\begin{equation}
l_{3}=a_{+}^{+}a_{+}-a_{-}^{+}a_{-},\label{}
\end{equation}
and
\begin{equation}
a_{1}^{+}a_{1}+a_{2}^{+}a_{2}=a_{+}^{+}a_{+}+a_{-}^{+}a_{-}.\label{}
\end{equation}
Therefore, we finally have
\begin{equation}
H_{\perp}^{2}=2a_{-}^{+}a_{-}+1-\sigma^{3}=\Bigg(\begin{array}{cc}
2a_{-}^{+}a_{-} & 0\\
0 & 2(a_{-}^{+}a_{-}+1)\end{array}\Bigg).\label{Hperp2}
\end{equation}
Here $a_{-}^{+}a_{-}$ is the boson number operators, and its
eigenvalue can be denoted as $n$ (n=0,1,2...). Recovering $eB$ we
obtain the eigenvalue of Eq.~\eqref{H2} with $\lambda^{2}=2neB$. We
should emphasize that from Eq.~\eqref{Hperp2} one can find that
states of $n>0$ are degenerate with two different spins, while for
$n=0$ state, there is only one spin. Therefore, this means that only
particles in the lowest level are polarized by the external magnetic
field.

Until now, we have solved the Dirac equation under a background
magnetic field. We summarize the results here. The wave function of
the fermion is given as
\begin{equation}
u^{s}=\Bigg(\begin{array}{c}\sqrt{E-\epsilon}\xi^{s}\\
\sqrt{E+\epsilon}\xi^{s}\end{array}\Bigg), \quad\mathrm{with}\quad
s=\pm 1.\label{}
\end{equation}
When $n>0$,
\begin{equation}
\xi^{s}=\frac{1}{\sqrt{2|\epsilon|}}\Bigg(\begin{array}{c}(s)\mathrm{sgn}(p_{3})\sqrt{|\epsilon|+s\lambda}\\
\sqrt{|\epsilon|-s\lambda}\end{array}\Bigg),\quad\mathrm{with}\quad
\epsilon=s|\epsilon|;\label{}
\end{equation}
when $n=0$,
\begin{equation}
\xi^{1}=\Bigg\{\begin{array}{c}(1,0)^{T}\quad \mathrm{for} \quad
p_{3}>0\\(0,1)^{T}\quad \mathrm{for} \quad p_{3}<0
\end{array},\quad\mathrm{with}\quad \epsilon=|p_{3}|,
\end{equation}
\begin{equation}
\xi^{-1}=\Bigg\{\begin{array}{c}(0,1)^{T}\quad \mathrm{for} \quad
p_{3}>0\\(1,0)^{T}\quad \mathrm{for} \quad p_{3}<0
\end{array},\quad\mathrm{with}\quad \epsilon=-|p_{3}|.
\end{equation}
The wave function of the anti-fermion is
\begin{equation}
v^{s}=\Bigg(\begin{array}{c}\sqrt{E-\epsilon}\xi^{-(s)}\\
-\sqrt{E+\epsilon}\xi^{-(s)}\end{array}\Bigg),
\quad\mathrm{with}\quad s=\pm 1.\label{}
\end{equation}
When $n>0$,
\begin{equation}
\xi^{-(s)}=\frac{1}{\sqrt{2|\epsilon|}}\Bigg(\begin{array}{c}(-s)\mathrm{sgn}(p_{3})\sqrt{|\epsilon|-s\lambda}\\
\sqrt{|\epsilon|+s\lambda}\end{array}\Bigg),\quad\mathrm{with}\quad
\epsilon=-s|\epsilon|;\label{}
\end{equation}
when $n=0$,
\begin{equation}
\xi^{-(1)}=\Bigg\{\begin{array}{c}(0,1)^{T}\quad \mathrm{for} \quad
p_{3}>0\\(1,0)^{T}\quad \mathrm{for} \quad p_{3}<0
\end{array},\quad\mathrm{with}\quad \epsilon=-|p_{3}|,
\end{equation}
\begin{equation}
\xi^{-(-1)}=\Bigg\{\begin{array}{c}(1,0)^{T}\quad \mathrm{for} \quad
p_{3}>0\\(0,1)^{T}\quad \mathrm{for} \quad p_{3}<0
\end{array},\quad\mathrm{with}\quad \epsilon=|p_{3}|.
\end{equation}
In the equations above, we have
\begin{eqnarray}
\lambda&=&\sqrt{2neB},\\
|\epsilon|&=&\sqrt{\lambda^{2}+p_{3}^{2}}=\sqrt{2neB+p_{3}^{2}},\\
E&=&\sqrt{|\epsilon|^{2}+m^{2}}=\sqrt{2neB+p_{3}^{2}+m^{2}}.
\end{eqnarray}
It should be emphasized that $s=+1$ represents the state that the
spin of a fermion or anti-fermion parallels to its momentum, i.e.
the right-handed state, while $s=-1$ corresponds to the state that
the spin anti-parallels to the momentum and thus is the left-handed
state. Therefore, the helicity of a fermion is $s$ and that of a
anti-fermion is $-s$.

Employing the standard canonical quantization procedure, we can
express the Hamiltonian $H$ in Eq.~\eqref{Hamiltonian} as
\begin{eqnarray}
\hat{H}&=&V\frac{e B}{2\pi}\sum_{n=0}^{\infty}\sum_{s=\pm
1}\int\frac{d
p_{3}}{2\pi}E({a_{\epsilon}^{s}}^{+}a_{\epsilon}^{s}-b_{\epsilon}^{s}{b_{\epsilon}^{s}}^{+})\nonumber\\
&=&V\frac{e B}{2\pi}\sum_{n=0}^{\infty}\sum_{s=\pm 1}\int\frac{d
p_{3}}{2\pi}E({a_{\epsilon}^{s}}^{+}a_{\epsilon}^{s}+{b_{\epsilon}^{s}}^{+}b_{\epsilon}^{s}-1),\label{pH}
\end{eqnarray}
where $a_{\epsilon}^{s}$ and $b_{\epsilon}^{s}$ correspond to the
annihilation operators of the fermion and anti-fermion,
respectively. $V$ is the volume of the system. Furthermore, in the
presence of an external magnetic field, we have
\begin{equation}
\int\int\frac{d p_{1}}{2\pi}\frac{d
p_{2}}{2\pi}\longrightarrow\frac{e B}{2\pi}\sum_{n=0}^{\infty}.
\end{equation}
In the same way, we have
\begin{eqnarray}
\hat{N}&=&\int d^{3}x\bar{\psi}\gamma^{0}\psi\nonumber\\
&=&V\frac{e B}{2\pi}\sum_{n=0}^{\infty}\sum_{s=\pm 1}\int\frac{d
p_{3}}{2\pi}\Big({a_{\epsilon}^{s}}^{+}a_{\epsilon}^{s}-{b_{\epsilon}^{s}}^{+}b_{\epsilon}^{s}\Big),\label{pN}
\end{eqnarray}
and
\begin{eqnarray}
\hat{N_{5}}&=&\int d^{3}x\bar{\psi}\gamma^{0}\gamma^{5}\psi\nonumber\\
&=&V\frac{e B}{2\pi}\sum_{n=0}^{\infty}\sum_{s=\pm 1}\int\frac{d
p_{3}}{2\pi}\Big(\frac{s|\epsilon|}{E}{a_{\epsilon}^{s}}^{+}a_{\epsilon}^{s}+\frac{s|\epsilon|}{E}{b_{\epsilon}^{s}}^{+}b_{\epsilon}^{s}\Big),\label{pN5}
\end{eqnarray}
Substituting Eqs.~\eqref{pH}~\eqref{pN}~\eqref{pN5} into
Eq.~\eqref{partition_function}, after a simple calculation we obtain
\begin{eqnarray}
\ln Z&=&V\frac{e B}{2\pi}\sum_{n=0}^{\infty}\sum_{s=\pm
1}\int\frac{d
p_{3}}{2\pi}\bigg(\frac{E}{T}+\ln\Big\{1+\exp\big[-(E-\mu\nonumber\\
&&-\frac{s|\epsilon|}{E}\mu_{5})/T\big]\Big\}+\ln\Big\{1+\exp\big[-(E+\mu
-\frac{s|\epsilon|}{E}\mu_{5})/T\big]\Big\}\bigg).\label{}
\end{eqnarray}

\section{Another approach with modified Lagrangian}
\setcounter{section}{2} \setcounter{equation}{0} \label{appendixB}

In the Appendix~\ref{appendixA}, we have derived the partition
function of a fermion system with $\mathcal{P}$ violation and under
a background magnetic field. We will discuss this subject in another
approach in this appendix.

Absorbing the term including $N_{5}$ in
Eq.~\eqref{partition_function} into the Hamiltonian $H$, We can
obtain the modified Lagrangian density
\begin{equation}
\mathcal{L}=\bar{\psi}(i\gamma^{\mu}D_{\mu}-m+\mu_{5}\gamma^{0}\gamma^{5})\psi.\label{}
\end{equation}
Then the Dirac equation in Eq.~\eqref{Dirac_equation} is also
modified as
\begin{equation}
i\partial_{0}\psi=\Big[(-i\partial_{i}-eA^{i})\gamma^{0}\gamma^{i}+\gamma^{0}m-\gamma^{5}\mu_{5}\Big]\psi.\label{}
\end{equation}
This Dirac equation can also be solved through the same method used
in the above appendix, and here we just give the results. The wave
function of the fermion is given as
\begin{equation}
u^{s}=\Bigg(\begin{array}{c}\sqrt{E-(\epsilon-\mu_{5})}\xi^{s}\\
\sqrt{E+(\epsilon-\mu_{5})}\xi^{s}\end{array}\Bigg),
\quad\mathrm{with}\quad s=\pm 1.\label{}
\end{equation}
When $n>0$,
\begin{equation}
\xi^{s}=\frac{1}{\sqrt{2|\epsilon|}}\Bigg(\begin{array}{c}(s)\mathrm{sgn}(p_{3})\sqrt{|\epsilon|+s\lambda}\\
\sqrt{|\epsilon|-s\lambda}\end{array}\Bigg),\quad\mathrm{with}\quad
\epsilon=s|\epsilon|;\label{}
\end{equation}
when $n=0$,
\begin{equation}
\xi^{1}=\Bigg\{\begin{array}{c}(1,0)^{T}\quad \mathrm{for} \quad
p_{3}>0\\(0,1)^{T}\quad \mathrm{for} \quad p_{3}<0
\end{array},\quad\mathrm{with}\quad \epsilon=|p_{3}|,
\end{equation}
\begin{equation}
\xi^{-1}=\Bigg\{\begin{array}{c}(0,1)^{T}\quad \mathrm{for} \quad
p_{3}>0\\(1,0)^{T}\quad \mathrm{for} \quad p_{3}<0
\end{array},\quad\mathrm{with}\quad \epsilon=-|p_{3}|.
\end{equation}
The wave function of the anti-fermion is
\begin{equation}
v^{s}=\Bigg(\begin{array}{c}\sqrt{E-(\epsilon+\mu_{5})}\xi^{-(s)}\\
-\sqrt{E+(\epsilon+\mu_{5})}\xi^{-(s)}\end{array}\Bigg),
\quad\mathrm{with}\quad s=\pm 1.\label{}
\end{equation}
When $n>0$,
\begin{equation}
\xi^{-(s)}=\frac{1}{\sqrt{2|\epsilon|}}\Bigg(\begin{array}{c}(-s)\mathrm{sgn}(p_{3})\sqrt{|\epsilon|-s\lambda}\\
\sqrt{|\epsilon|+s\lambda}\end{array}\Bigg),\quad\mathrm{with}\quad
\epsilon=-s|\epsilon|;\label{}
\end{equation}
when $n=0$,
\begin{equation}
\xi^{-(1)}=\Bigg\{\begin{array}{c}(0,1)^{T}\quad \mathrm{for} \quad
p_{3}>0\\(1,0)^{T}\quad \mathrm{for} \quad p_{3}<0
\end{array},\quad\mathrm{with}\quad \epsilon=-|p_{3}|,
\end{equation}
\begin{equation}
\xi^{-(-1)}=\Bigg\{\begin{array}{c}(1,0)^{T}\quad \mathrm{for} \quad
p_{3}>0\\(0,1)^{T}\quad \mathrm{for} \quad p_{3}<0
\end{array},\quad\mathrm{with}\quad \epsilon=|p_{3}|.
\end{equation}
In the equations above, we have
\begin{eqnarray}
\lambda&=&\sqrt{2neB},\\
|\epsilon|&=&\sqrt{\lambda^{2}+p_{3}^{2}}=\sqrt{2neB+p_{3}^{2}},\\
E&=&\sqrt{(|\epsilon|-s\mu_{5})^{2}+m^{2}}.\label{energy}
\end{eqnarray}
The partition function is given by
\begin{eqnarray}
\ln Z&=&V\frac{e B}{2\pi}\sum_{n=0}^{\infty}\sum_{s=\pm
1}\int\frac{d
p_{3}}{2\pi}\bigg(\frac{E}{T}+\ln\Big\{1+\exp\big[-(E-\mu\nonumber)/T\big]\Big\}\\
&&+\ln\Big\{1+\exp\big[-(E+\mu)/T\big]\Big\}\bigg).\label{lnZ}
\end{eqnarray}
We should emphasize that the expression of particle energy $E$ in
the above equation is given by Eq.~\eqref{energy}.

In the same way, we can obtain
\begin{eqnarray}
\hat{N_{5}}&=&\int d^{3}x\bar{\psi}\gamma^{0}\gamma^{5}\psi\nonumber\\
&=&V\frac{e B}{2\pi}\sum_{n=0}^{\infty}\sum_{s=\pm 1}\int\frac{d
p_{3}}{2\pi}\Big[\frac{(s|\epsilon|-\mu_{5})}{E}{a_{\epsilon}^{s}}^{+}a_{\epsilon}^{s}\nonumber\\
&&+\frac{(s|\epsilon|-\mu_{5})}{E}{b_{\epsilon}^{s}}^{+}b_{\epsilon}^{s}-\frac{(s|\epsilon|-\mu_{5})}{E}\Big].\label{}
\end{eqnarray}
Making the ensemble average of $\hat{N_{5}}$ we have
\begin{eqnarray}
N_{5}&\equiv&\langle\hat{N_{5}}\rangle\nonumber\\
&=&V\frac{e B}{2\pi}\sum_{n=0}^{\infty}\sum_{s=\pm 1}\int\frac{d
p_{3}}{2\pi}\Big[\frac{(s|\epsilon|-\mu_{5})}{E}\frac{1}{e^{(E-\mu)/T}+1}\nonumber\\
&&+\frac{(s|\epsilon|-\mu_{5})}{E}\frac{1}{e^{(E+\mu)/T}+1}-\frac{(s|\epsilon|-\mu_{5})}{E}\Big].\label{}
\end{eqnarray}
$N_{5}$ can also been directly obtained through $N_{5}=T\partial \ln
Z/\partial \mu_{5}$.

Next, we calculate the chiral electric current
density~\cite{Fukushima2008}:
\begin{eqnarray}
\hat{j_{3}}&=&\frac{e}{V}\int d^{3}x\bar{\psi}\gamma^{3}\psi\nonumber\\
&=&\frac{e^{2} B}{2\pi}\sum_{n=0}^{\infty}\sum_{s=\pm 1}\int\frac{d
p_{3}}{2\pi}\Big[\big(\frac{s|\epsilon|-\mu_{5}}{E}\big)\big(\frac{sp_{3}}{|\epsilon|}\big){a_{\epsilon}^{s}}^{+}a_{\epsilon}^{s}\nonumber\\
&&-\big(\frac{s|\epsilon|-\mu_{5}}{E}\big)\big(\frac{sp_{3}}{|\epsilon|}\big){b_{\epsilon}^{s}}^{+}b_{\epsilon}^{s}
+\big(\frac{s|\epsilon|-\mu_{5}}{E}\big)\big(\frac{sp_{3}}{|\epsilon|}\big)\Big].\label{j3o}
\end{eqnarray}
Therefore, the ensemble average of $\hat{j_{3}}$ is
\begin{eqnarray}
j_{3}&\equiv&\langle\hat{j_{3}}\rangle\nonumber\\
&=&\frac{e^{2} B}{2\pi}\sum_{n=0}^{\infty}\sum_{s=\pm 1}\int\frac{d
p_{3}}{2\pi}\Big[\big(\frac{s|\epsilon|-\mu_{5}}{E}\big)\big(\frac{sp_{3}}{|\epsilon|}\big)\frac{1}{e^{(E-\mu)/T}+1}\nonumber\\
&&-\big(\frac{s|\epsilon|-\mu_{5}}{E}\big)\big(\frac{sp_{3}}{|\epsilon|}\big)\frac{1}{e^{(E+\mu)/T}+1}
+\big(\frac{s|\epsilon|-\mu_{5}}{E}\big)\big(\frac{sp_{3}}{|\epsilon|}\big)\Big].\label{j3}
\end{eqnarray}
Same as $N_{5}$, $j_{3}$ can also been obtained through
\begin{equation}
j_{3}=-\frac{eT}{V}\Big(\frac{\partial \ln Z^{+}}{\partial
p_{3}}-\frac{\partial \ln Z^{-}}{\partial p_{3}}\Big),\label{j3Z}
\end{equation}
where
\begin{equation}
\ln Z^{+}=V\frac{e B}{2\pi}\sum_{n=0}^{\infty}\sum_{s=\pm
1}\int\frac{d
p_{3}}{2\pi}\ln\Big\{1+\exp\big[-(E-\mu\nonumber)/T\big]\Big\},\label{}
\end{equation}
and
\begin{equation}
\ln Z^{-}=V\frac{e B}{2\pi}\sum_{n=0}^{\infty}\sum_{s=\pm
1}\int\frac{d
p_{3}}{2\pi}\bigg(\frac{E}{T}+\ln\Big\{1+\exp\big[-(E+\mu)/T\big]\Big\}\bigg),\label{}
\end{equation}
are the fermion part and anti-fermion part of the partition function
in Eq.~\eqref{lnZ}, respectively. The minus related with the
anti-fermion part in Eq.~\eqref{j3Z} is due to our convention in
Eq.~\eqref{Diracmin}.

From Eq.~\eqref{j3} we can find that only the states with $n=0$
contribute to the chiral electric current density $j_{3}$, since the
integral variable $p_{3}$ ranges from $-\infty$ to $\infty$ for
$n>0$, while from $-\infty$ to $0$ or $0$ to $\infty$ for $n=0$.
Therefor, $j_{3}$ can be simplified as
\begin{eqnarray}
j_{3} &=&\frac{e^{2} B}{2\pi}\sum_{s=\pm 1}\int\frac{d
p_{3}}{2\pi}\Big[\big(\frac{s|\epsilon|-\mu_{5}}{E}\big)\big(\frac{sp_{3}}{|\epsilon|}\big)\frac{1}{e^{(E-\mu)/T}+1}\nonumber\\
&&-\big(\frac{s|\epsilon|-\mu_{5}}{E}\big)\big(\frac{sp_{3}}{|\epsilon|}\big)\frac{1}{e^{(E+\mu)/T}+1}
+\big(\frac{s|\epsilon|-\mu_{5}}{E}\big)\big(\frac{sp_{3}}{|\epsilon|}\big)\Big].\label{j30}
\end{eqnarray}
In order to understand the physical meanings of the several terms in
Eq.~\eqref{j30}, we just extract the term related with the positive
fermion and calculate
\begin{eqnarray}
&&\sum_{s=\pm 1}\int d p_{3}\big(\frac{s|\epsilon|-\mu_{5}}{E}\big)\big(\frac{sp_{3}}{|\epsilon|}\big)\frac{1}{e^{(E-\mu)/T}+1}\nonumber\\
&=&\int_{0}^{\infty}d
p_{3}\frac{p_{3}-\mu_{5}}{\sqrt{(p_{3}-\mu_{5})^{2}+m^{2}}}\frac{1}{\exp[(\sqrt{(p_{3}-\mu_{5})^{2}+m^{2}}
-\mu)/T]+1}\nonumber\\
&&+\int_{-\infty}^{0}d
p_{3}\frac{p_{3}-\mu_{5}}{\sqrt{(p_{3}-\mu_{5})^{2}+m^{2}}}\frac{1}{\exp[(\sqrt{(p_{3}-\mu_{5})^{2}+m^{2}}
-\mu)/T]+1}\nonumber\\
&=&0.\label{}
\end{eqnarray}

From the above equation we can find that, in the modified Lagrangian
approach, it is $p_{3}-\mu_{5}$ not $p_{3}$ that judges whether a
particle is right-handed or left-handed, i.e., if the sign of
$p_{3}-\mu_{5}$ is same as that of the particle spin along the
$z$-direction, the particle is right-hand; if opposite, then the
particle is left-hand. However, the direction of motion of the
particle is governed by $p_{3}$ not $p_{3}-\mu_{5}$. Therefore, the
properties of right-handed or left-handed of a particle cannot
determine the direction of motion of the particle. In another word,
the left-right asymmetry cannot be correctly related with charge
separation effect in this approach. So, we conclude that the
modified Lagrangian approach is inappropriate to describe the chiral
magnetic effect. We should point out that the approach given in
Appendix~\ref{appendixA} does not have this problem.

Next, we continue to calculate the chiral electric current density
$j_{3}$ in Eq.~\eqref{j30}. The only nonvanishing contribution to
$j_{3}$ come from the last term in Eq.~\eqref{j30}, i.e.
\begin{eqnarray}
j_{3} &=&\frac{e^{2} B}{(2\pi)^{2}}\sum_{s=\pm 1}\int d
p_{3}\big(\frac{s|\epsilon|-\mu_{5}}{E}\big)\big(\frac{sp_{3}}{|\epsilon|}\big)\nonumber\\
&=&\frac{e^{2} B}{(2\pi)^{2}}\int_{-\infty}^{\infty}d
p_{3}\frac{p_{3}+\mu_{5}}{\sqrt{(p_{3}+\mu_{5})^{2}+m^{2}}}\nonumber\\
&=&\frac{e^{2} B}{(2\pi)^{2}}\int_{-\Lambda}^{\Lambda}d
p_{3}\frac{p_{3}+\mu_{5}}{\sqrt{(p_{3}+\mu_{5})^{2}+m^{2}}}\Big|_{\lim
\Lambda\rightarrow\infty}\nonumber\\
&=&\frac{e^{2} B\mu_{5}}{2\pi^{2}}\label{j3vac}
\end{eqnarray}
This is the result obtained in Ref.~\cite{Fukushima2008}. From our
calculations above, one can find that $j_{3}$ calculated in the
modified Lagrangian approach does not comes from the finite
temperature contributions, however its source is the Dirac Sea which
can be clearly seen in Eq.~\eqref{j3o}. Therefore, it is argued in
Ref.~\cite{Fukushima2008} that $j_{3}$ is independent of the
temperature, chemical potential $\mu$, and even the mass of the
particle (from the calculations above we find that the independence
of $j_{3}$ on the mass of the particle is due to the fact that the
ultraviolet momentum in the Dirac Sea makes the mass of particle
meaningless). We should comment that the results obtained in the
modified Lagrangian approach are contrary to our physical intuition.
On the one hand, the physical observable (here is the $j_{3}$) is
unlikely to come from the Dirac sea. On the other hand, the chiral
electric current density is also unlikely to be independent of the
properties of current carriers. In this work we will show that the
chiral electric density $j_{3}$ calculated in an appropriate
approach, i.e., the approach in Appendix~\ref{appendixA}, comes from
the finite temperature contributions not the Dirac Sea, and $j_{3}$
is also dependent of the properties of current carriers and is also
influenced by the external environment.

\end{document}